\RequirePackage[2020-02-02]{latexrelease}

\documentclass[%
 reprint,
 amsmath,amssymb,
 aps,
]{revtex4-2}
\usepackage{float}
\usepackage{graphicx}
\usepackage{dcolumn}
\usepackage{bm}
\usepackage{subcaption,lipsum}
\usepackage[margin=1.5cm]{geometry}

\usepackage{mathrsfs}

\begin{document}

\preprint{APS/123-QED}

\title{Unified Spectroscopic Study of Bottom Mesons and Doubly Bottom Baryons Using a Relativistic Flux-Tube Model}
\author{Pooja Jakhad}
\author{Ajay Kumar Rai}
\affiliation{Department of Physics, Sardar Vallabhbhai National Institute of Technology, Surat, Gujarat-395007, India}
\date{\today}
\begin{abstract}
	Using the relativistic flux tube (RFT) model with spin-dependent corrections, we present a unified investigation of the mass spectra for singly bottom mesons ($B$, $B_s$) and doubly bottom baryons ($\Xi_{bb}$, $\Omega_{bb}$). Within this framework, the calculated masses for established $B$ and $B_s$ meson states show excellent agreement with experimental data. Leveraging these reliable predictions for low-lying states, we propose spectroscopic assignments for several higher excitations: we identify the $B_J^*(5732)$ resonance as an excellent candidate for the $1P(1^+)$ state; further, we assign the $B_J(5840)$ resonance as the first radial excitation ($2S$), and the $B_J(5970)$ resonance as a $1D$-wave orbital excitation. In the bottom-strange sector, we predict that the recently observed $B_{sJ}(6063)$ and $B_{sJ}(6114)$ resonances belong to the $1D$-wave multiplet. Extending our model to doubly bottom baryons, we present comprehensive predictions for the spectra of the $\Xi_{bb}$ and $\Omega_{bb}$ baryons, identifying their ground and excited states. These theoretical insights serve as valuable benchmarks, offering  guidance for future experimental searches targeting yet-to-be-discovered bottom mesons and doubly bottom baryons.
\end{abstract}
\maketitle


\section{INTRODUCTION}

The spectroscopy of open-bottom hadrons – encompassing bottom mesons ($B$ and $B_s$) and doubly bottom baryons ($\Xi_{bb}$ and $\Omega_{bb}$) – provides a crucial window into the nonperturbative dynamics of quantum chromodynamics (QCD). These systems involve one or more heavy $b$ quarks bound to light quark constituents, situating them at the interface of heavy-quark physics and light-quark confinement. Because the $b$ quark mass is large (of order 4–5 GeV), $B$ and $B_s$ hadrons serve as exemplary testing grounds for heavy quark symmetry and the interplay between the heavy quark and the QCD confining potential. Precise knowledge of their mass spectra is not only important for understanding hadronic structure, but also for refining theoretical models of the strong interaction in the heavy-quark sector.


Over the past two decades, significant experimental progress has been made in the bottom meson sector. The ground-state $B(1S)$ mesons ($B^{0,\pm}$ and $B^{*}$), as well as the $B_s(1S)$ mesons ( $B_s^{0}$ and $B_s^*$)  were established long ago and are well documented in the Particle Data Group (PDG) listings \cite{PDG2024}. In 2007, the first orbitally excited bottom mesons were discovered: the narrow $P$-wave states $B_{1}(5721)^{0,+}$ and $B_{2}^{*}(5747)^{0,+}$ (with $J^P=1^+$ and $2^+$) were observed by D0 Collaboration \cite{D0:2007vzd} and quickly confirmed by CDF Collaboration \cite{CDF:2008qzb}, and their strange counterparts $B_{s1}(5830)$ and $B_{s2}^{*}(5840)$ were reported by CDF Collaboration \cite{CDF:2007avt} in the same year and subsequently confirmed by D0 and LHCb Collaborations \cite{LHCb:2012iuq,D0:2007die}. 

In 2013, CDF Collaboration announced the observation of a higher-mass resonance $B(5970)$ in the $B\pi$ channel \cite{CDF:2013www}. LHCb later resolved two resonant structures labeled $B_J(5840)$ and $B_J(5960)$ in proton–proton collision data \cite{LHCb:2015aaf}, with properties of $B_J(5960)$  consistent with the earlier $B(5970)$ signal observed by CDF Collaboration. More recently, the LHCb Collaboration reported evidence for two excited $B_s$ mesons ${B_{sJ}(6064)}$ and ${B_{sJ}(6114)}$ in the $B^+K^-$ invariant mass spectrum \cite{LHCb:2020pet}. This indicates that as experimental sensitivity has improved, higher excited bottom meson candidates have begun to appear. However, the quantum numbers of some of these higher states remain undetermined. Clarifying the nature of such excitations is an ongoing challenge. In this context, a detailed theoretical spectrum of $B$ and $B_s$ mesons is valuable to interpret new resonances and to guide searches for missing states.

On the baryon side, the situation is more challenging. The $\Xi_{bb}$ and $\Omega_{bb}$ baryons, containing two $b$ quarks, have not yet been observed. Nevertheless, a major precedent was set by the discovery of the doubly charmed baryon $\Xi_{cc}^{++}$ in 2017, first observed by LHCb \cite{LHCb:2017iph}. This marked the first confirmed instance of a baryon with two heavy quarks that provides strong motivation that the bottom–bottom counterparts are real and could be found given sufficient data and refined techniques. Knowing their masses within a reasonable range is crucial for guiding experimental searches and optimizing selection criteria.

The progress on the experimental side has been paralleled by extensive theoretical efforts to calculate and classify the spectra of $B$ and $B_s$ mesons and doubly bottom baryons. A variety of theoretical methods, including the relativistic quark model \cite{Ebert:2009ua, Godfrey:2016nwn, Sun:2014wea, li:2021hss, Zeng:1994vj, Shah:2016mgq, Devlani:2011zz, Lu:2017meb,Ebert:2002ig,Gershtein:2000nx,Yoshida:2015tia,Giannuzzi:2009gh}, non-relativistic quark model \cite{li:2021hss, Rai:2008sc, Kher:2017mky, Yazarloo:2016luc, Patel:2024cng, Eakins:2012jk, PhysRevD.105.074014, Kiselev:2002iy, Roberts:2007ni, Shah:2016vmd, Shah:2017liu, Karliner:2014gca}, $^{3}P_{0}$ model \cite{Lu:2017hma}, Regge phenomenology \cite{Oudichhya:2024hmn}, heavy quark effective theory \cite{Korner:1994nh}, heavy meson chiral perturbation theory \cite{Cheng:2017oqh, Alhakami:2020vil}, QCD sum rules \cite{Bagan:1992za, Wang:2010hs}, the union of heavy quark effective theory and chiral perturbation theory \cite{Alhakami:2020vil}, and the
lattice QCD \cite{Lang:2015hza, Gregory:2010gm, Padmanath:2015jea, Alexandrou:2012xk}.

Although numerous theoretical and experimental efforts have been made, the quark model classification of several excited states—such as $B_J(5840)$, $B_J(5970)$, $B_{sJ}(6063)$, and $B_{sJ}(6114)$—remains ambiguous. Moreover, the ground states of doubly bottom baryons have not yet been experimentally confirmed. As the theoretical approaches mentioned so far  offer a different perspective on the mass spectrum, further theoretical studies are essential. In the current study, we employ the relativistic flux-tube derived mass formula to compute spin-averaged masses. Crucially, our methodology integrates spin-dependent interactions, enabling predictions of hyperfine-structure splittings within a $jj$ coupling scheme suitable for heavy-light systems. We aim to identify the masses of fairly high orbital and radial excited states. This approach has been validated in our previous research, successfully reproducing mass spectra for singly heavy baryons \cite{Jakhad:2023ids,Jakhad:2024fin,Jakhad:2024wpx}, open-charm mesons and doubly charm baryons \cite{Private_communication}. In the present work, we extend the application of this model to open-bottom mesons and doubly bottom baryons. Based on the predicted mass spectra, we further propose quark model classification for several experimentally observed charmed meson states whose quantum numbers remain unconfirmed. Additionally, we provide theoretical predictions for the excitation spectrum of the doubly bottom baryons $\Xi_{bb}$ and $\Omega_{bb}$, which may serve as valuable input for future experimental searches aimed at discovering and classifying their excited states.

The structure of the paper is organized as follows. In Section II, we outline the relativistic flux-tube model as applied to bottom mesons and doubly bottom baryons, and detail the methodology used to compute their mass spectra, including the treatment of spin-dependent interactions. In Section III, we present our computed spectra for $B$, $B_s$, $\Xi_{bb}$, and $\Omega_{bb}$, and compare these results with existing theoretical predictions and available experimental data. Section IV contains our conclusions and outlook, summarizing the main findings.

\section{THEORETICAL FRAMEWORK}

We employ the relativistic flux tube (RFT) model to describe both the bottom mesons ($B, B_s$) and the doubly bottom baryons ($\Xi_{bb}, \Omega_{bb}$) in a unified two-body picture. In the case of a meson, the two constituents are the bottom quark ($b/ \bar{b}$) and a light quark ($q / \bar q$), while for a doubly bottom baryon the constituents are a tightly-bound $\{bb\}$ diquark and a light quark ($q=u, d$ for $\Xi_{bb}$ or $s$ for $\Omega_{bb}$). These two constituents are connected by a flux tube – a tube like structure containing gluonic field - with constant tension $T$ that provides the confining potential. The heavy and light constituents bound by the flux tube, rotate around their common center of mass with quantized orbital angular momentum. 
The light component is highly relativistic (effectively moving at velocity $v\approx c$), whereas the heavy component moves more slowly due to its large mass.  In this limit, the flux-tube model yields a linear Regge-like relation between the mass ($\bar{M}$) and the orbital angular momentum quantum number $L$. Specifically, neglecting spin interactions, the spin-averaged mass $\bar M$ satisfies (for both mesons and baryons) a relation of the form \cite{article,Jakhad:2023ids}:
\begin{equation}
	\label{eq1}
	(\bar{M} - m_h)^2 = \sigma\,\frac{L}{2} \;+\; (m_l + m_h v_h^2)^2.
\end{equation}
Here,  $\sigma = 2\pi T$. The factor $v_h$ is the speed of the heavy component introduced to account for relativistic effects. Similarly, the light component has velocity $v_l$. The dynamical masses for the rotating  light component (having mass $ m_{0_l}$) and heavy component (having  mass $ m_{0_h}$) are $ m_l = m_{0_l}/\sqrt{1-v_l^2}$ and $ m_h = m_{0_h}/\sqrt{1-v_h^2}$, respectively. 

The distance between the heavy and light constituents in this model is given as \cite{article, Jakhad:2023ids}
\begin{equation}
	\label{eq2}
	r =  (v_l + v_h)\sqrt{\frac{8L}{\sigma}}.
\end{equation}

In the framework of a two-body heavy–light hadronic system, the quantum solution of the relativistic flux-tube model predicts that the Regge trajectories in the $(L, (\bar{M} - m_h)^2)$ plane—corresponding to various radial excitations—are both parallel and equidistant \cite{Olson:1993ux,Olsson:1993cn,Allen:1999dk}. This observation necessitates a refinement of the semi classical expressions (\ref{eq1}) and (\ref{eq2}) to include the radial excitations, wherein $L$ is replaced with $\lambda n_r + L$. Here, $n_r = n - 1$, with $n$ denoting the principal quantum number (i.e., $n = 1, 2, 3, \ldots$) that indexes the radial excitation states. The resulting relationships are expressed as follows \cite{article,Jakhad:2023ids}:
\begin{equation}
	\label{eq3}
	(\bar{M} - m_h)^2 = \sigma\,\frac{\big[L + \lambda\,n_r\big]}{2} \;+\; (m_l + m_h v_h^2)^2
\end{equation}
\begin{equation}
	\label{eq4}
	r =  (v_l + v_h)\sqrt{\frac{8\big[L + \lambda\,n_r\big]}{\sigma}}.
\end{equation}
In this formulation, $\lambda$ serves as a model parameter that quantifies the vertical separation between Regge trajectories corresponding to different principal quantum numbers $n = 1, 2, 3, \ldots$ in the $(L, (\bar{M} - m_h)^2)$ plane.

Having obtained the spin-averaged mass $\bar M$ of each level, we now include spin-dependent interactions. For a heavy–light meson and baryon, the total angular momentum $\mathbf{J}$ arises from coupling the orbital angular momentum $\mathbf{L}$ with the spins of the heavy and light constituents ($\mathbf{S_h}$ and $\mathbf{S_l}$, respectively). There are two possible coupling schemes to couple $\mathbf{S_h}$, $\mathbf{S_l}$ and $\mathbf{L}$: $L$–$S$ coupling, where $\mathbf{S_h}$ and $\mathbf{S_l}$ are first combined into a total spin $\mathbf{S} = \mathbf{S_h} + \mathbf{S_l}$ which is then coupled to $\mathbf{L}$ to give $\mathbf{J}$; and $j$–$j$ coupling, where one first adds the light spin to the orbital momentum ($\mathbf{j} = \mathbf{L} + \mathbf{S_l}$ is the total angular momentum of the light degrees of freedom) and then couples $\mathbf{j}$ with the heavy spin $\mathbf{S_h}$ to obtain $\mathbf{J}$. Owing to heavy quark symmetry, the $j$–$j$ coupling scheme is the natural choice for heavy–light systems. We therefore adopt the $j$–$j$ coupling scheme for both $B/B_s$ mesons (where spin quantum number of heavy constituent is $S_h = \tfrac{1}{2}$ for the $b$ quark) and doubly bottom baryons (where spin quantum number of heavy constituent  $S_h$ is the total angular momentum of the $\{bb\}$ diquark. In this scheme, each orbital excitation $L$ splits into multiplet states characterized by the quantum numbers $j$ and $J$. The splittings around the spin-averaged mass $\bar M$ are governed by spin-dependent interactions.

We include the standard spin–orbit,  spin–spin, and tensor interaction terms as a perturbation to the flux-tube spectrum. The Hamiltonian for spin-dependent effects can be written in the form                                                                                                                         
\begin{equation}\label{eq2'}
	\Delta M \;=\; H_{so} \;+\; H_{t} \;+\; H_{ss}\,.
\end{equation}
The spin–orbit interaction term, $H_{so}$, originates from the combined effects of short-range color magnetic interactions and long-range Thomas precession. It is expressed as a coupling between the orbital angular momentum $\mathbf{L}$ and the spin operators of the light and heavy constituents, $\mathbf{S}_l$ and $\mathbf{S}_h$, respectively \cite{PhysRevD.105.074014}:

\begin{equation}\label{eq2.1}
	\begin{aligned}
		H_{so} &= \Big[\left( \frac{2\alpha}{3r^3} - \frac{b_0}{2r} \right)\frac{1}{m_l^2}
		+ \frac{4\alpha}{3r^3}\frac{1}{m_h m_l}\Big]\mathbf{L}\!\cdot\!\mathbf{S}_l \\
		&\quad+ \Big[\left( \frac{2\alpha}{3r^3} - \frac{b_0}{2r} \right)\frac{1}{m_h^2}
		+ \frac{4\alpha}{3r^3}\frac{1}{m_h m_l}\Big]\mathbf{L}\!\cdot\!\mathbf{S}_h
	\end{aligned}
\end{equation}

where $\alpha$ is the strong coupling constant, and $b_0$ characterizes the long-range confining interaction. The tensor interaction term, $H_t$, arises from the long-range component of the color-magnetic dipole–dipole interaction. It is represented as \cite{PhysRevD.105.074014}
\begin{equation}\label{eq2.2}
	H_t = \frac{4\alpha}{3 r^{3} m_h m_l}\,\hat{\mathbf{B}},
\end{equation}
with the tensor operator $\hat{\mathbf{B}}$ defined as:
$
\hat{\mathbf{B}} = {3(\mathbf{S}_l \cdot \mathbf{r})(\mathbf{S}_h \cdot \mathbf{r})}/{r^{2}} - \mathbf{S}_l \cdot \mathbf{S}_h.
$
The spin–spin contact interaction, $H_{ss}$, corresponds to a short-range hyperfine term and is written as \cite{PhysRevD.105.074014}:

\begin{equation}\label{eq2.3}
	H_{ss} = \frac{32\pi \alpha \sigma_0^3}{9\sqrt{\pi} \, m_h m_l}\, e^{-\sigma_{0}^{2} r^{2}} \mathbf{S}_l \cdot \mathbf{S}_h,
\end{equation}
where $\sigma_0$ denoting the width parameter.

We have evaluated the expectation values of the operators $\mathbf{L} \cdot \mathbf{S}_l$, $\mathbf{L} \cdot \mathbf{S}_h$, $\mathbf{S}_l \cdot \mathbf{S}_h$, and $\hat{\mathbf{B}}$ in the $]J,j\rangle$ basis, and present the results in Table \ref{table0.1} and \ref{table0.2}. The detailed calculation for these expectation values for the $S$-, $P$-, and $D$-wave states have been computed in our earlier work \cite{Private_communication}, while those corresponding to the $F$- and $G$-wave states are detailed in the Appendix. Finally, the physical mass of the state is obtained by adding this spin correction to the spin-averaged mass: $\bar{M} +\Delta M $. In this way, the relativistic flux tube model provides a complete framework to calculate the mass spectrum of heavy–light mesons ($B$ and $B_s$) and doubly heavy baryons ($\Xi_{bb}$ and $\Omega_{bb}$).

\begin{table}
	\caption{\label{table0.1}
		Expectation values of the spin–orbit operators $\mathbf{L\cdot S_{l}}$ and $\mathbf{L\cdot S_{h}}$, the tensor operator $\mathbf{\hat{B}}$, and the contact hyperfine term $\mathbf{S_{h} \cdot S_{l}}$ in the $|J,j\rangle$ basis for heavy–light mesons.  Results are listed for every allowed $(L,J,j)$ configuration (up to $G$-wave). \cite{Private_communication}.}
	\begin{ruledtabular}
		\begin{tabular}{ccccccc}
			\multicolumn{3}{c}{Quantum Numbers} &  &  &  &  \\
			\cline{1-3}
			$L$ & $j$ &$J$ &  $\;\langle\mathbf{L}\!\cdot\!\mathbf{S}_l\rangle\;$ & $\;\langle\mathbf{L}\!\cdot\!\mathbf{S}_h\rangle\;$ & $\;\langle\mathbf{\hat{B}}\rangle\;$ & $\;\langle\mathbf{S}_h\!\cdot\!\mathbf{S}_l\rangle\;$ \\
			\hline
			0 & 1/2  & 0   &   0   &    0    &   0   &  –3/4  \\
			0 & 1/2  & 1   &   0   &    0    &   0   &   1/4  \\
			1 & 1/2  & 0   &  –1   &   –1    &  –1   &   1/4  \\
			1 & 1/2  & 1   &  –1   &  1/3    &  1/3  & –1/12  \\
			1 & 3/2  & 1   &  1/2  &  –5/6   &  1/6  & –5/12  \\
			1 & 3/2  & 2   &  1/2  &   1/2   & –1/10 &   1/4  \\
			2 & 3/2  & 1   & –3/2  &  –3/2   & –1/2  &   1/4  \\
			2 & 3/2  & 2   & –3/2  &   9/10  &  3/10 & –3/20  \\
			2 & 5/2  & 2   &   1   &  –7/5   &  1/5  & –7/20  \\
			2 & 5/2  & 3   &   1   &    1    & –1/7  &   1/4  \\
			3 & 5/2  & 2   &  –2   &   –2    & –2/5  &   1/4  \\
			3 & 5/2  & 3   &  –2   &  10/7   &  2/7  & –5/28  \\
			3 & 7/2  & 3   &  3/2  & –27/14  &  3/14 & –9/28  \\
			3 & 5/2  & 4   &  3/2  &   3/2   & –1/6  &   1/4  \\
			4 & 7/2  & 3   & –5/2  &  –5/2   & –5/14 &   1/4  \\
			4 & 7/2  & 4   & –5/2  &  35/18  &  5/18 & –7/36  \\
			4 & 9/2  & 4   &   2   &  –22/9  &  2/9  & –11/36 \\
			4 & 9/2  & 5   &   2   &    2    & –2/11 &   1/4  \\
			5 & 9/2  & 4   &  –3   &   –3    & –1/3  &   1/4  \\
			5 & 9/2  & 5   &  –3   &  27/11  &  3/11 & –9/44  \\
			5 & 11/2 & 5   &  5/2  & –65/22  &  5/22 & –13/44 \\
			5 & 11/2 & 6   &  5/2  &   5/2   & –5/26 &   1/4  \\
		\end{tabular}
	\end{ruledtabular}
\end{table}

\begin{table}
	\caption{\label{table0.2} Same set of expectation values as in Table \ref{table0.1}, but for doubly bottom baryons treated in the quark–diquark picture with axial-vector diquark ($S_h=1$) \cite{Private_communication}.}
	\begin{ruledtabular}
		\begin{tabular}{ccccccc}
			\multicolumn{3}{c}{Quantum Numbers} &  &  &  &  \\
			\cline{1-3}
			$L$ & $j$ & $J$ &  $\;\langle\mathbf{L}\!\cdot\!\mathbf{S}_l\rangle\;$ & $\;\langle\mathbf{L}\!\cdot\!\mathbf{S}_h\rangle\;$ & $\;\langle\mathbf{\hat{B}}\rangle\;$ & $\;\langle\mathbf{S}_h\!\cdot\!\mathbf{S}_l\rangle\;$ \\
			\hline
			0 & 1/2 & 1/2  &   0   &    0   &    0   &  –1    \\
			0 & 1/2 & 3/2  &   0   &    0   &    0   &  1/2   \\
			1 & 1/2 & 1/2  &  –1   & –4/3   & –4/3   &  1/3   \\
			1 & 3/2 & 1/2  &  1/2  & –5/3   &  1/3   & –5/6   \\
			1 & 1/2 & 3/2  &  –1   &  2/3   &  2/3   & –1/6   \\
			1 & 3/2 & 3/2  &  1/2  & –2/3   &  2/15  & –1/3   \\
			1 & 3/2 & 5/2  &  1/2  &   1    & –1/5   &  1/2   \\
			2 & 3/2 & 1/2  & –3/2  &   –3   & –1     &  1/2   \\
			2 & 3/2 & 3/2  & –3/2  &  –6/5  & –2/5   &  1/5   \\
			2 & 5/2 & 3/2  &   1   & –14/5  &  2/5   & –7/10  \\
			2 & 3/2 & 5/2  & –3/2  &   9/5  &  3/5   & –3/10  \\
			2 & 5/2 & 5/2  &   1   &  –4/5  &  4/35  & –1/5   \\
			2 & 5/2 & 7/2  &   1   &    2   & –2/7   &  1/2   \\
			3 & 5/2 & 1/2  &  –2   &   –4   & –4/5   &  1/2   \\
			3 & 5/2 & 5/2  &  –2   &  –8/7  & –8/35  &  1/7   \\
			3 & 7/2 & 5/2  &  3/2  & –27/7  &  3/7   & –9/14  \\
			3 & 5/2 & 7/2  &  –2   &  20/7  &  4/7   & –5/14  \\
			3 & 7/2 & 7/2  &  3/2  &  –6/7  &  2/21  & –1/7   \\
			3 & 7/2 & 9/2  &  3/2  &    3   & –1/3   &  1/2   \\
			4 & 7/2 & 5/2  & –5/2  &   –5   & –5/7   &  1/2   \\
			4 & 7/2 & 7/2  & –5/2  & –70/63 & –10/63 &  7/63  \\
			4 & 9/2 & 7/2  &   2   & –44/9  &  4/9   & –11/18 \\
			4 & 7/2 & 9/2  & –5/2  &  35/9  &  5/9   & –7/18  \\
			4 & 9/2 & 9/2  &   2   &  –8/9  &  8/99  & –1/9   \\
			4 & 9/2 & 11/2 &   2   &    4   & –4/11  &  1/2   \\
		\end{tabular}
	\end{ruledtabular}
\end{table} 

Now we proceed to discussion on selection and determination of the model parameters for the system of heavy-light mesons and doubly heavy baryons.
\subsubsection{Heavy-light mesons}

We employed the iminuit Python package \cite{iminuit2021}, which implements the robust MINUIT minimization algorithm developed by CERN \cite{minuit1975}, to determine optimal values and associated uncertainties of our model parameters for the heavy-light mesons ($D$, $D_s$, $B$, $B_s$). The input data for this optimization procedure included experimentally measured masses, along with their uncertainties, for well established mesonic states \cite{PDG2024}: $D^{\pm}$, $D^{0}$, $D^*(2007)^0$, $D^*(2010)^\pm$, $D_0^{*}(2300)$, $D_1(2420)$, $D_1(2430)^0$, $D_2^{*}(2460)$, $D_3^{*}(2750)$, $D_s^{\pm}$, $D_s^{*\pm}$, $D_{s1}(2536)^{\pm}$, $D_{s2}^{*}(2573)$, $D_{s1}^{*}(2700)^{\pm}$, $D_{s3}^{*}(2860)^{\pm}$, $B^{\pm, 0}$, $B^{*}$, $B_{1}(5721)$, $B_{2}^{*}(5747)$, $B_{s}^{0}$, $B_{s}^{*}$, $B_{s1}(5830)^{0}$, $B_{s2}^{*}(5840)^{0}$, mass of charm quark $m_{0_{c}}=1273\pm4.6~\text{MeV}$ and mass of bottom quark $m_{0_{b}}=4183\pm7~\text{MeV}$. The best-fit parameter values obtained, along with their statistical uncertainties, are as follows: the dynamical mass of the charm quark $m_c = 1375.68 \pm 0.15$ MeV, the bottom quark $m_b =4431.08 \pm 0.16$ MeV, the up/down quark $m_{u/d} = 399.95 \pm 0.02$ MeV, and the strange quark $m_s = 484.29 \pm 0.06$ MeV. The string tensions determined for various mesons are: $\sigma_D = (1.5321 \pm 0.0023) \times 10^6$ MeV$^2$, $\sigma_{D_{s}} = (1.7882 \pm 0.0016) \times 10^{6}~~\text{MeV}^2$, $\sigma_B = (1.7541 \pm 0.0031) \times 10^6~~\text{MeV}^2$, and $\sigma_{B_{s}} = (1.9931 \pm 0.0006) \times 10^{6}~~\text{MeV}^2$. Additional model parameters include $\lambda = 1.479 \pm 0.015$, $b_{0} = (36.2 \pm 1.2) \times 10^3~~\text{MeV}^2$,  $\sigma_0 = 502.8 \pm 2.8$ MeV, and the coupling constant for $B$ and $B_s$ mesons $\alpha = 0.317 \pm 0.0004$. Specifically, from the $B$ and $B_s$ meson sectors, experimentally measured masses of the $1S(0^-,1/2)$, $1S(1^-,1/2)$, $1P(1^+,3/2)$, and $1P(2^+,3/2)$ states are used for fitting the model parameters. With these fitted parameters the masses of all higher excited $B$ and $B_s$  mesonic states can be predicted.

\subsubsection{Doubly heavy baryons with the diquark in the ground state}

{Due to the Pauli exclusion principle, the diquark in the ground state (in internal $1S$-wave) within doubly bottom baryon must form an axial-vector diquark, denoted as $\{bb\}$ with spin $S_h = 1$.}

For doubly bottom baryons ($\Xi_{bb}$ and $\Omega_{bb}$), we set the model parameters $m_{u/d}$, $m_s$, $\lambda$, $b_{0}$, and $\sigma_0$ identical to those for the $B$ and $B_s$ mesons. The mass of the heavy diquark ($\{bb\}$) in its internal $1S$ wave is approximated as the sum of individual bottom quark masses, yielding $m_{0_{\{bb\}}} = 2m_{0_{b}} = 8366$ MeV \cite{PDG2024}. Due to the scarcity of experimental measurements for doubly heavy baryons, except for the $\Xi_{cc}^{++}$, it is essential to leverage theoretical predictions for $1S$-wave states to constrain our parameters. A notable theoretical prediction by Ebert et al. \cite{Ebert:2002ig}, estimating the $\Xi_{cc}$ ground state mass at $3620$ MeV, closely matched the subsequent experimental discovery by the LHCb Collaboration at $3621.55$ MeV \cite{LHCb:2019epo}. Encouraged by this precision, we adopted theoretical predictions from Ref. \cite{Ebert:2002ig} for the ground-state masses of $\Xi_{bb}$ and $\Omega_{bb}$ baryons, enabling us to estimate the dynamical diquark mass as $m_{\{bb\}}= 8861.7$ MeV and coupling constant as $\alpha = 0.324$. Using the  mass of the ($\{bb\}$)  diquark in its internal $1S$ wave $m_{0_{\{bb\}}} =  8366$ MeV and the dynamical mass of the diquark $m_{\{bb\}}= 8861.7$ MeV, the speed of the heavy diquark is extracted to be the $v_h=0.33 \pm 0.0022$.  

Using the mass of the diquark $\{bb\}$ in its internal $1S$-wave, 
$m_{0_{\{bb\}}} = 8366 \pm 8 ~\text{MeV}$, and the dynamical mass of the diquark, 
$m_{\{bb\}} = 8861.7\pm 0.000006 ~\text{MeV}$, the speed of the heavy diquark is obtained as 
$v_{h} = 0.33 \pm 0.0022$.

In the heavy quark limit ($m_{h}\rightarrow\infty$), two heavy quarks forming a diquark inside a baryon behave similarly to a heavy antiquark inside a meson \cite{PhysRevD.73.054003}. Consequently, the light quark in a doubly heavy baryon experiences the same color field environment as in heavy-light mesons. Heavy-quark diquark symmetry, therefore, implies identical string tensions at leading order for such systems. However, when considering finite quark masses, slight deviations from exact symmetry occur. These finite mass effects can be effectively captured by empirical power-law scaling relations connecting string tensions and heavy quark masses \cite{jia2023}:

\begin{equation} \label{x}
	\frac{\sigma_{B}}{\sigma_{D}} = \left(\frac{m_b}{m_c}\right)^P, \quad
	\frac{\sigma_{B}}{\sigma_{\Xi_{bb}}} = \left(\frac{m_b}{m_{\{bb\}}}\right)^P.
\end{equation}
Analogously, for strange heavy-light systems, the scaling relations are given by \cite{jia2023}:

\begin{equation}\label{y}
	\frac{\sigma_{B_{s}}}{\sigma_{D_{s}}} = \left(\frac{m_b}{m_c}\right)^Q, \quad
	\frac{\sigma_{B_{s}}}{\sigma_{\Omega_{bb}}} = \left(\frac{m_b}{m_{\{bb\}}}\right)^Q.
\end{equation}

Using the fitted parameters $\sigma_{D}$, $\sigma_{D_{s}}$, $\sigma_{B}$, $\sigma_{B_{s}}$, $m_c$, and $m_{b}$ in Eqs.~[\ref{x}, \ref{y}] the scaling parameters were extracted as $P = 0.1157 \pm 0.002$ and $Q = 0.0927 \pm 0.0017$. Substituting these values back into in Eqs.~[\ref{x}, \ref{y}], the string tensions for the baryonic systems were calculated as $\sigma_{\Xi_{bb}} = (1.901 \pm 0.008) \times 10^6$ MeV$^2$ and $\sigma_{\Omega_{bb}} = (2.125 \pm 0.007) \times 10^6$ MeV$^2$. So, the model parameters determined from the mesonic sector can facilitate reliable mass predictions for  $\Xi_{bb}$, and $\Omega_{bb}$ baryons.

\subsubsection{Doubly heavy baryons with the diquark in an excited state}

We now consider configurations in which the two bottom quarks are excited relative to each other inside the diquark. In the internal excitations, the \(bb\) diquark must be in a spin–singlet configuration (\(S_h=0\)) for the \(1P\) and \(2P\) waves, and in a spin–triplet configuration (\(S_h=1\)) for the \(2S\) wave.

We can calculate the masses for $\Xi_{bb}$ and $\Omega_{bb}$ states where the light quark is excited orbitally or radially with respect to the heavy diquark in the internal  $2S$,  $1P$ , and  $2P$ waves using Eqs. [\ref{eq3}, \ref{eq2.1}-\ref{eq2.3}] To apply these relations we first calculate the mass of diquarks  $m_{h_{2S}}$, $m_{h_{1P}}$, and $m_{h_{2P}}$  in the internal $2S$,  $1P$ , and  $2P$ waves, respectively. 

To this end, we first  consider the bottomonium in relativistic flux tube model. A bottom quark and anti bottom quark of mass $m_b$ connected by a flux tube and carrying relative orbital angular momentum $L$ and principle quantum number $n=n_r+1$ obey a Regge-like relation for the spin-averaged mass $\bar M$   \cite{PhysRevD.101.014020}:
\begin{equation}
	\label{eq3.1}
	\bar{M} = 2m_b +  \left( \frac{\sigma_{b\bar{b}}^2}{2 \pi^{2} m_b} \right)^{\tfrac{1}{3}}(\lambda n_r + L)^{\frac{2}{3}},
\end{equation}
where $\sigma$ is the string tension for bottomonium. Using this relation, the relation between the spin average mass for $1S$ and $1P$ waves is
\begin{equation}
	\bar{M}_{1P}= \bar{M}_{1S} + \left( \frac{\sigma_{b\bar{b}}^2}{\pi^{2}\bar{M}_{1S}} \right)^{\tfrac{1}{3}}
\end{equation}
We use the measured masses of $1S$ and $1P$ waves of bottomonium, named $\eta_{b}(1S)$, $\Upsilon(1S)$, $\chi_{b0}(1P)$, $\chi_{b1}(1P)$, $h_{b}(1P)$, and $\chi_{b2}(1P)$,  to calculate the spin average masses of $1S$ and $1P$ waves. With that we extract $\sigma_{b\bar{b}}=(2.961 \pm 0.006) \times 10^6$ MeV$^2 $. 

Building upon this framework for bottomonium, we then extend Eq. [\ref{eq3.1}] to the $bb$ diquark, specified by the principal quantum number $n_d=n_{rd}+1$ and orbital angular momentum quantum number $L_d$ as

\begin{equation}
	\label{eq3.2}
	\bar{M} = 2m_b +\left( \frac{\sigma_{bb}^2}{2 \pi^{2}m_b} \right)^{\tfrac{1}{3}}(\lambda n_{rd} + L_d)^{\frac{2}{3}},
\end{equation}
This leads to the following relations for the spin–averaged  mass of the diquark in $1P$, $2S$ and $2P$ waves: 
\begin{equation}
	\label{x1}
	\bar{M}_{1P} = \bar{M}_{1S} +  \left( \frac{\sigma_{bb}^2}{\pi^{2}\bar{M}_{1S}} \right)^{\tfrac{1}{3}}
\end{equation}
\begin{equation}
	\label{x2}
	\bar{M}_{2S} = \bar{M}_{1S} + \left( \frac{\sigma_{bb}^2}{\pi^{2}\bar{M}_{1S}} \right)^{\tfrac{1}{3}} \lambda^{\frac{2}{3}}
\end{equation}
\begin{equation}
	\label{x3}
	\bar{M}_{2P} = \bar{M}_{1S} +  \left( \frac{\sigma_{bb}^2}{\pi^{2}\bar{M}_{1S}} \right)^{\tfrac{1}{3}}(\lambda+1)^{\frac{2}{3}}
\end{equation}
The inverse Regge slope $1/ \alpha'$ (and hence the flux-tube string tension $\sigma$, via $1/\alpha'\!\sim\! \sigma$) proportional to the square-root of the color factor: $1/\alpha' \propto \sqrt{C\,\alpha_s}$, where $C$ is the quadratic Casimir of the color sources at the ends of the tube \cite{PhysRevD.13.1934}. For a bottom diquark in the antitriplet, $bb[\bar{\mathbf 3}_c]$, one has $C=2/3$; for a bottom quark–antiquark color singlet, $b\bar b[\mathbf 1_c]$, $C=4/3$. Consequently, the corresponding string tensions satisfy 
\begin{equation}
	\sigma_{\,bb}=\sqrt{\frac{2/3}{4/3}}\sigma_{\,b\bar b}
	=\frac{1}{\sqrt{2}}\sigma_{\,b\bar b}
\end{equation}
With this relation, and using the value of $\sigma_{\,b\bar b}$, the string tension between the two bottom quarks in the $bb$ diquark is obtained as $\sigma_{\,bb}=(2.094 \pm 0.004) \times 10^6$ MeV$^2$.
Employing this value of $\sigma_{\,bb}$ together with the previously determined $\bar{M}_{1S}=8366 \pm 8$~MeV and $\lambda = 1.479 \pm 0.015$, we extract the spin-averaged masses of the excited diquark states using Eq. [\ref{x1}-\ref{x3}] as $\bar{M}_{2S}=8854 \pm 9$ MeV, $\bar{M}_{1P}=8742 \pm 8$ MeV, and $\bar{M}_{2P}=9054 \pm 8$ MeV.

These excited diquarks are considered in either a spin-singlet configuration ($S_h = 0$) in the internal $1P$ and $2P$ waves, or a spin-triplet configuration ($S_h = 1$) in the internal $2S$ wave, where they are connected to the light quark via a flux tube, and the entire system revolves around its center of mass.
The dynamical masses of the rotating diquarks, with velocity $v_{h} = 0.33 \pm 0.0022$, are extracted as $m_{h_{2S}} = 9379 \pm 12\ \text{MeV}$, $m_{h_{1P}} = 9261 \pm 11\ \text{MeV}$, and $m_{h_{2P}} = 9592 \pm 12\ \text{MeV}$. Using these dynamical masses of the diquarks in excited states, we calculate the spectra of the $\Xi_{bb}$ and $\Omega_{bb}$ baryons, where the light quark is excited either orbitally or radially with respect to the heavy diquark in the internal $2S$, $1P$, and $2P$ waves. These calculations are performed through Eqs. [\ref{eq3}, \ref{eq2.1}-\ref{eq2.3}], employing the previously fitted parameters $m_{u/d}$, $m_s$, $\lambda$, $b_{0}$, $\sigma_0$, $v_h$, $\sigma_{\Xi_{bb}}$, and $\sigma_{\Omega_{bb}}$.

\begin{widetext}
	\begin{table*}
		\caption{\label{table1}	Masses of $B$-Meson (in MeV).}
		
		\begin{ruledtabular}
			\begin{tabular}{ccccccc}
				$n L (J^{P}, j) $   &     MASS      &     PDG \cite{PDG2024}     & \cite{Ebert:2009ua} & \cite{Godfrey:2016nwn} & \cite{Sun:2014wea} & \cite{li:2021hss} \\ \hline
				$1 S (0^{-}, 1/2) $  & 5279  $\pm$   15 &     5279.57  $\pm$  0 .05     & 5280                & 5312                   & 5309               & 5279              \\
				$1 S (1^{-}, 1/2) $  & 5325  $\pm$   15 &     5324.75  $\pm$  0 .20     & 5326                & 5371                   & 5369               & 5325              \\
				$2 S (0^{-}, 1/2) $  & 5870  $\pm$   10 &                            & 5890                & 5904                   & 5904               & 5876              \\
				$2 S (1^{-}, 1/2) $  & 5872  $\pm$   10 &                            & 5906                & 5933                   & 5934               & 5899              \\
				$1 P (0^{+}, 1/2) $  & 5670  $\pm$   11 &                            & 5749                & 5756                   & 5796               & 5722              \\
				$1 P (1^{+}, 1/2) $  & 5696  $\pm$   11 & $5698 \pm 8$                           & 5774                & 5777                   & 5756               & 5716              \\
				$1 P (1^{+}, 3/2) $  & 5722  $\pm$   10 & 5726.0 $^{+2.5} _ {-2.7} $ & 5723                & 5784                   & 5782               & 5753              \\
				$1 P (2^{+}, 3/2) $  & 5738  $\pm$   10 &      5737.3  $\pm$  0 .7      & 5741                & 5797                   & 5779               & 5727              \\
				$2 P (0^{+}, 1/2) $  & 6152  $\pm$   9  &                            & 6221                & 6213                   & 6213               &                   \\
				$2 P (1^{+}, 1/2) $  & 6159  $\pm$   9  &                            & 6281                & 6197                   & 6214               &                   \\
				$2 P (1^{+}, 3/2) $  & 6144  $\pm$   8  &                            & 6209                & 6228                   & 6219               &                   \\
				$2 P (2^{+}, 3/2) $  & 6147  $\pm$   8  &                            & 6260                & 6213                   & 6206               &                   \\
				$1 D (1^{-}, 3/2) $  & 6022  $\pm$   9  &                            & 6119                & 6110                   & 6105               & 6056              \\
				$1 D (2^{-}, 3/2) $  & 6033  $\pm$   8  &                            & 6121                & 6095                   & 6110               & 5973              \\
				$1 D (2^{-}, 5/2) $  & 6014  $\pm$   8  &                            & 6103                & 6124                   & 6113               & 6067              \\
				$1 D (3^{-}, 5/2) $  & 6022  $\pm$   8  &                            & 6091                & 6106                   & 6108               & 5979              \\
				$2 D (1^{-}, 3/2) $  & 6401  $\pm$   8  &                            & 6534                & 6475                   & 6459               &                   \\
				$2 D (2^{-}, 3/2) $  & 6406  $\pm$   8  &                            & 6554                & 6450                   & 6475               &                   \\
				$2 D (2^{-}, 5/2) $  & 6375  $\pm$   8  &                            & 6528                & 6486                   & 6472               &                   \\
				$2 D (3^{-}, 5/2) $  & 6378  $\pm$   8  &                            & 6542                & 6460                   & 6464               &                   \\
				$1 F (2^{+}, 5/2) $  & 6293  $\pm$   7  &                            & 6412                & 6387                   & 6364               &                   \\
				$1 F (3^{+}, 5/2) $  & 6301  $\pm$   7  &                            & 6420                & 6358                   & 6387               &                   \\
				$1 F (3^{+}, 7/2) $  & 6259  $\pm$   7  &                            & 6391                & 6396                   & 6380               &                   \\
				$1 F (4^{+}, 7/2) $  & 6265  $\pm$   7  &                            & 6380                & 6364                   & 6375               &                   \\
				$2 F (2^{+}, 5/2) $  & 6621  $\pm$   7  &                            &                     &                        &                    &                   \\
				$2 F (3^{+}, 5/2) $  & 6626  $\pm$   7  &                            &                     &                        &                    &                   \\
				$2 F (3^{+}, 7/2) $  & 6581  $\pm$   7  &                            &                     &                        &                    &                   \\
				$2 F (4^{+}, 7/2) $  & 6584  $\pm$   7  &                            &                     &                        &                    &                   \\
				$1 G (3^{-}, 7/2) $  & 6527  $\pm$   7  &                            & 6664                & 6622                   &                    &                   \\
				$1 G (4^{-}, 7/2) $  & 6533  $\pm$   7  &                            & 6652                & 6588                   &                    &                   \\
				$1 G (4^{-}, 9/2) $  & 6476  $\pm$   7  &                            & 6648                & 6628                   &                    &                   \\
				$1 G (5^{-}, 9/2) $  & 6481  $\pm$   6  &                            & 6634                & 6592                   &                    &                   \\
				$1 H (4^{+}, 9/2) $  & 6736  $\pm$   6  &                            &                     &                        &                    &                   \\
				$1 H (5^{+}, 9/2) $  & 6741  $\pm$   6  &                            &                     &                        &                    &                   \\
				$1 H (5^{+}, 11/2) $ & 6672  $\pm$   6  &                            &                     &                        &                    &                   \\
				$1 H (6^{+}, 11/2) $ & 6676  $\pm$   6  &                            &                     &                        &                    &                   		\end{tabular}
		\end{ruledtabular}
	\end{table*}
	
	\begin{table*}
		\caption{\label{table2}	Masses of $B_s$-Meson
			(in MeV).}

		\begin{ruledtabular}
			
			\begin{tabular}{ccccccc}
				$n L (J^{P}, j) $   &      MASS       & PDG \cite{PDG2024} & \cite{Ebert:2009ua} & \cite{Godfrey:2016nwn} & \cite{Sun:2014wea} & \cite{li:2021hss} \\ \hline
				$1 S (0^{-}, 1/2) $  & 5369  $\pm$     15 &  5366.93 $\pm$  0.1   & 5372                & 5394                   & 5390               & 5367              \\
				$1 S (1^{-}, 1/2) $  & 5407  $\pm$     15 &   5415.4 $\pm$  1.4   & 5414                & 5450                   & 5447               & 5415              \\
				$2 S (0^{-}, 1/2) $  &  5981	 $\pm$  	10  &                    & 5976                & 5984                   & 5985               & 5944              \\
				$2 S (1^{-}, 1/2) $  &  5984	 $\pm$  	10  &                    & 5992                & 6012                   & 6013               & 5966              \\
				$1 P (0^{+}, 1/2) $  &  5772	 $\pm$  	11  &                    & 5833                & 5831                   & 5875               & 5788              \\
				$1 P (1^{+}, 1/2) $  &  5798	 $\pm$  	11  &                    & 5865                & 5857                   & 5830               & 5810              \\
				$1 P (1^{+}, 3/2) $  &  5825	 $\pm$  	10  &  5828.73 $\pm$  0.2   & 5831                & 5861                   & 5859               & 5821              \\
				$1 P (2^{+}, 3/2) $  &  5841	 $\pm$  	10  &  5839.88 $\pm$  0.12  & 5842                & 5876                   & 5858               & 5821              \\
				$2 P (0^{+}, 1/2) $  &  6275	 $\pm$  	9   &                    & 6318                & 6279                   & 6295               &                   \\
				$2 P (1^{+}, 1/2) $  &  6282	 $\pm$  	9   &                    & 6345                & 6279                   & 6279               &                   \\
				$2 P (1^{+}, 3/2) $  &  6273	 $\pm$  	9   &                    & 6321                & 6296                   & 6291               &                   \\
				$2 P (2^{+}, 3/2) $  &  6276	 $\pm$  	9   &                    & 6359                & 6295                   & 6284               &                   \\
				$1 D (1^{-}, 3/2) $  &  6135	 $\pm$  	9   &                    & 6209                & 6182                   & 6178               & 6101              \\
				$1 D (2^{-}, 3/2) $  &  6146	 $\pm$  	9   &                    & 6218                & 6169                   & 6181               & 6061              \\
				$1 D (2^{-}, 5/2) $  &  6137	 $\pm$  	8   &                    & 6189                & 6196                   & 6185               & 6113              \\
				$1 D (3^{-}, 5/2) $  &  6145	 $\pm$  	8   &                    & 6191                & 6179                   & 6180               & 6067              \\
				$2 D (1^{-}, 3/2) $  &  6536	 $\pm$  	8   &                    & 6629                & 6542                   & 6534               &                   \\
				$2 D (2^{-}, 3/2) $  &  6541	 $\pm$  	8   &                    & 6651                & 6526                   & 6542               &                   \\
				$2 D (2^{-}, 5/2) $  &  6521	 $\pm$  	8   &                    & 6625                & 6553                   & 6542               &                   \\
				$2 D (3^{-}, 5/2) $  &  6524	 $\pm$  	8   &                    & 6637                & 6535                   & 6536               &                   \\
				$1 F (2^{+}, 5/2) $  &  6418	 $\pm$  	7   &                    & 6501                & 6454                   & 6431               &                   \\
				$1 F (3^{+}, 5/2) $  &  6426	 $\pm$  	7   &                    & 6515                & 6425                   & 6453               &                   \\
				$1 F (3^{+}, 7/2) $  &  6400	 $\pm$  	7   &                    & 6468                & 6462                   & 6446               &                   \\
				$1 F (4^{+}, 7/2) $  &  6406	 $\pm$  	7   &                    & 6475                & 6432                   & 6441               &                   \\
				$2 F (2^{+}, 5/2) $  &  6767	 $\pm$  	7   &                    &                     &                        &                    &                   \\
				$2 F (3^{+}, 5/2) $  &  6771	 $\pm$  	7   &                    &                     &                        &                    &                   \\
				$2 F (3^{+}, 7/2) $  &  6741	 $\pm$  	7   &                    &                     &                        &                    &                   \\
				$2 F (4^{+}, 7/2) $  &  6744	 $\pm$  	7   &                    &                     &                        &                    &                   \\
				$1 G (3^{-}, 7/2) $  &  6664	 $\pm$  	7   &                    & 6753                & 6685                   &                    &                   \\
				$1 G (4^{-}, 7/2) $  &  6670	 $\pm$  	7   &                    & 6762                & 6650                   &                    &                   \\
				$1 G (4^{-}, 9/2) $  &  6632	 $\pm$  	6   &                    & 6715                & 6690                   &                    &                   \\
				$1 G (5^{-}, 9/2) $  &  6637	 $\pm$  	6   &                    & 6726                & 6654                   &                    &                   \\
				$1 H (4^{+}, 9/2) $  &  6884	 $\pm$  	6   &                    &                     &                        &                    &                   \\
				$1 H (5^{+}, 9/2) $  &  6889	 $\pm$  	6   &                    &                     &                        &                    &                   \\
				$1 H (5^{+}, 11/2) $ &  6842	 $\pm$  	6   &                    &                     &                        &                    &                   \\
				$1 H (6^{+}, 11/2) $ &  6846	 $\pm$  	6   &                    &                     &                        &                    &                   		\end{tabular}
		\end{ruledtabular}
	\end{table*}

	\begin{table*}
		\caption{\label{table3.1} Masses of $\Xi_{bb}$-baryon (in MeV).}
		\begin{ruledtabular}
			\begin{tabular}{ccllllccccccc}
				$n_dL_dnL(J^{P},j)$    &       MASS        & Ref.\cite{Oudichhya:2024hmn} & Ref.\cite{PhysRevD.105.074014} &  Ref.\cite{Yoshida:2015tia} & Ref.\cite{jia2023} & Ref.\cite{Lu:2017meb} & Ref.\cite{Eakins:2012jk} & Ref.\cite{Ebert:2002ig} & Ref.\cite{Gershtein:2000nx} & Ref. \cite{Yoshida:2015tia} & \cite{Giannuzzi:2009gh} &  \\ \hline
				$1S1s(1/2^{+},1/2)$    &  10203 $\pm$  29  & 10225                        & 10171                          &10202 &10314 &         10138         &          10322           &          10202          &            10093            &            10314            &          10185          &  \\
				$1S1s(3/2^{+},1/2)$    &  10238 $\pm$  29  & 10273                        & 10195                          &10237 &10339 &         10169         &          10352           &          10237          &            10133            &            10339            &          10216          &  \\
				$1S2s(1/2^{+},1/2)$    &  10668 $\pm$  23  & 10609                        & 10738                          &10707 &10612 &         10662         &          10940           &          10832          &                             &                             &          10751          &  \\
				$1S2s(3/2^{+},1/2)$    &  10671 $\pm$  22  & 10617                        & 10753                          &10726 &10593 &         10675         &          10972           &          10860          &                             &                             &          10770          &  \\

				$1S1p(1/2^{-},1/2)$    &  10493 $\pm$  25  & 10497                        & 10593                          & 10523& 10476&         10525         &          10694           &          10632          &            10541            &            10703            &                         &  \\
				$1S1p(1/2^{-},3/2)$    &  10544 $\pm$  24  & 10502                        & 10606                          &10559 & &         10526         &          10691           &          10647          &            10578            &            10704            &                         &  \\
				$1S1p(3/2^{-},1/2)$    &  10516 $\pm$  24  & 10494                        & 10547                          &10545 &10476 &         10504         &          10691           &          10675          &            10567            &            10740            &                         &  \\
				$1S1p(3/2^{-},3/2)$    &  10550 $\pm$  24  & 10506                        & 10561                          &10564 & &         10528         &          10692           &          10694          &            10581            &            10742            &                         &  \\
				$1S1p(5/2^{-},3/2)$    &  10561 $\pm$  23  & 10551                        & 10560                          &10574 &10759 &         10547         &          10695           &          10661          &            10580            &            10759            &                         &  \\
				$1P1s(1/2^{-}, 1/2)$    &  10647 $\pm$  19  &                              &                                &10608 & 10740&   10364                    &    10470                      &          10368          &                             &                             &                         &  \\
				$1P1s(3/2^{-}, 1/2)$    &  10681 $\pm$  19  &                              &                                &10640 &10742 &    10387                   & 10470                         &          10408          &                             &                             &                         &  \\
				$2S1s(1/2^{+}, 1/2)$   &  10777 $\pm$  34  &                              &                                &10550 & 10571&     10464                  &  10551                        &          10441          &                             &                             &       10453                  &  \\
				$2S1s(3/2^{+}, 1/2)$    &  10810 $\pm$  34  &                              &                                &10583 & 10592& 10480                      &   10574                      &          10482          &                             &                             &     10478                    &  \\	
				$1S2 p (1/2^{-}, 1/2) $  & 10920	 $\pm$  	20 &                              &                                & & &                       &                          &                         &                             &                             &                         &  \\
				$1S2 p (1/2^{-}, 3/2) $  & 10910	 $\pm$  	20 &                              &                                & & &                       &                          &                         &                             &                             &                         &  \\
				$1S2 p (3/2^{-}, 1/2) $  & 10926	 $\pm$  	20 & 10765                        &                                & & &                       &                          &                         &                             &                             &                         &  \\
				$1S2 p (3/2^{-}, 3/2) $  & 10911	 $\pm$  	20 &                              &                                & & &                       &                          &                         &                             &                             &                         &  \\
				$1S2 p (5/2^{-}, 3/2) $  & 10913	 $\pm$  	20 & 10776                        &                                & & &                       &                          &                         &                             &                             &                         &  \\ 
				$1S1d(1/2^{+},3/2)$    &  10799 $\pm$  21  &                              & 10913                          & & &                       &          11011           &                         &                             &                             &                         &  \\
				$1S1d(3/2^{+},3/2)$    &  10804 $\pm$  21  & 10699                        & 10918                          & & &                       &          11011           &                         &                             &                             &                         &  \\
				$1S1d(3/2^{+},5/2)$    &  10793 $\pm$  21  & 10742                        & 10921                          & & &                       &          11002           &                         &                             &                             &                         &  \\
				$1S1d(5/2^{+},3/2)$    &  10812 $\pm$  21  & 10756                        & 10798                          & & &                       &          11011           &                         &                             &                             &                         &  \\
				$1S1d(5/2^{+},5/2)$    &  10797 $\pm$  21  & 10708                        & 10803                          & & &                       &          11002           &                         &                             &                             &                         &  \\
				$1S1d(7/2^{+},5/2)$    &  10802 $\pm$  20  & 10823                        & 10805                          & & &                       &          11011           &                         &                             &                             &                         &  \\
				$2P1s(1/2^{-}, 1/2)$    &  11015 $\pm$  20  &                              &                                & & &                       &                          &          10563          &                             &                             &                         &  \\
				$2P1s(3/2^{-}, 1/2)$    &  11047 $\pm$  20  &                              &                                & & &                       &                          &          10607          &                             &                             &                         &   \\
				$1S2 d (1/2^{+}, 3/2) $  & 11148	 $\pm$  	18 &                              &                                & & &                       &                          &                         &                             &                             &                         &  \\
				$1S2 d (3/2^{+}, 3/2) $  & 11151	 $\pm$  	18 &                              &                                & & &                       &                          &                         &                             &                             &                         &  \\
				$1S2 d (3/2^{+}, 5/2) $  & 11122	 $\pm$  	18 &                              &                                & & &                       &                          &                         &                             &                             &                         &  \\
				$1S2 d (5/2^{+}, 3/2) $  & 11154	 $\pm$  	18 & 10901                        &                                & & &                       &                          &                         &                             &                             &                         &  \\
				$1S2 d (5/2^{+}, 5/2) $  & 11124	 $\pm$  	18 &                              &                                & & &                       &                          &                         &                             &                             &                         &  \\
				$1S2 d (7/2^{+}, 5/2) $  & 11126	 $\pm$  	18 &                              &                                & & &                       &                          &                         &                             &                             &                         &  \\
				$1S1 f (3/2^{-}, 5/2) $  & 11048	 $\pm$  	19 &                              &                                & & &                       &                          &                         &                             &                             &                         &  \\
				$1S1 f (5/2^{-}, 5/2) $  & 11051	 $\pm$  	19 & 10897                        &                                & & &                       &                          &                         &                             &                             &                         &  \\
				$1S1 f (5/2^{-}, 7/2) $  & 11014	 $\pm$  	19 & 10978                        &                                & & &                       &                          &                         &                             &                             &                         &  \\
				$1S1 f (7/2^{-}, 5/2) $  & 11057	 $\pm$  	19 & 11012                        &                                & & &                       &                          &                         &                             &                             &                         &  \\
				$1S1 f (7/2^{-}, 7/2) $  & 11017	 $\pm$  	19 & 10907                        &                                & & &                       &                          &                         &                             &                             &                         &  \\
				$1S1 f (9/2^{-}, 7/2) $  & 11021	 $\pm$  	19 & 11087                        &                                & & &                       &                          &                         &                             &                             &                         &  \\
				$1S1 g (5/2^{+}, 7/2) $  & 11267	 $\pm$  	17 &                              &                                & & &                       &                          &                         &                             &                             &                         &  \\
				$1S1 g (7/2^{+}, 7/2) $  & 11270	 $\pm$  	17 & 11092                        &                                & & &                       &                          &                         &                             &                             &                         &  \\
				$1S1 g (7/2^{+}, 9/2) $  & 11215	 $\pm$  	17 & 11208                        &                                & & &                       &                          &                         &                             &                             &                         &  \\
				$1S1 g (9/2^{+}, 7/2) $  & 11274	 $\pm$  	17 & 11262                        &                                & & &                       &                          &                         &                             &                             &                         &  \\
				$1S1 g (9/2^{+}, 9/2) $  & 11218	 $\pm$  	17 & 11102                        &                                & & &                       &                          &                         &                             &                             &                         &  \\
				$1S1 g (11/2^{+}, 9/2) $  & 11221	 $\pm$  	17 & 11346                        &                                & & &                       &                          &                         &                             &                             &                         &  \\
			\end{tabular}
		\end{ruledtabular}
	\end{table*}

	\begin{table*}
		\caption{\label{table4}	Masses of $\Omega_{bb}$-baryon (in MeV). }
		\begin{ruledtabular}
			\begin{tabular}{ccllllccccccc}
				$n_dL_dnL(J^{P},j)$        &       MASS        & Ref. \cite{Oudichhya:2024hmn} & Ref. \cite{PhysRevD.105.074014} & Ref. \cite{Yoshida:2015tia} &  Ref. \cite{jia2023} & Ref.\cite{Lu:2017meb} & Ref.\cite{Ebert:2002ig} & Ref.\cite{Kiselev:2002iy} & Ref. \cite{Yoshida:2015tia} & Ref.\cite{Roberts:2007ni} & Ref. \cite{Giannuzzi:2009gh} &  \\ \hline
				$1S1s(1/2^{+},1/2)$    &  10292 $\pm$  29  & 10350                    & 10266                      & &10447 &         10230         &          10359          &           10210           &            10447            &           10454           &          10271          &  \\
				$1S1s(3/2^{+},1/2)$    &  10321 $\pm$  29  & 10393                    & 10291                      & &10707 &         10258         &          10389          &           10257           &            10467            &           10486           &          10289          &  \\
				$1S2s(1/2^{+},1/2)$    &  10776 $\pm$  23  & 10736                    & 10816                      &10845 & 10744&         10751         &          10970          &                           &                             &                           &          10830          &  \\
				$1S2s(3/2^{+},1/2)$    &  10779 $\pm$  23  & 10743                    & 10830                      &10858 &10730 &         10763         &          10992          &                           &                             &                           &          10839          &  \\ 
				$1S1p(1/2^{-},1/2)$    &  10596 $\pm$  25  & 10622                    & 10669                      &10651 & 10607&         10605         &          10771          &           10541           &            10796            &                           &                         &  \\
				$1S1p(1/2^{-},3/2)$    &  10644 $\pm$  24  & 10643                    & 10681                      &10678 & 10796&         10610         &          10785          &           10567           &            10797            &                           &                         &  \\
				$1S1p(3/2^{-},1/2)$    &  10618 $\pm$  24  & 10633                    & 10641                      &10672 & 10608&         10591         &          10804          &           10578           &            10803            &           10763           &                         &  \\
				$1S1p(3/2^{-},3/2)$    &  10651 $\pm$  24  & 10647                    & 10656                      &10683 & 10797&         10611         &          10821          &           10581           &            10805            &           10765           &                         &  \\
				$1S1p(5/2^{-},3/2)$    &  10662 $\pm$  24  & 10667                    & 10655                      &10692 &10808 &         10625         &          10798          &           10580           &            10808            &           10766           &                         &  \\
				$1P1s(1/2^{-}, 1/2)$    &  10735 $\pm$  19  &                          &                            & 10768 & 10803&     10464                  &     10532                    &                           &                             &                           &                         &  \\
				$1P1s(3/2^{-}, 1/2)$    &  10763 $\pm$  19  &                          &                            & 10791 &10805 &      10482                 &     10566                    &                           &                             &                           &                         &  \\
				$2S1s(1/2^{+}, 1/2)$    &  10865 $\pm$  34  &                          &                            &10710 &10707 &     10566                  &    10610                     &                           &                             &                           &      10538                   &  \\
				$2S1s(3/2^{+}, 1/2)$    &  10893 $\pm$  34  &                          &                            & 10733& 10723&     10579                  &      10645                   &                           &                             &                           &   10556                      &  \\
				$1S2 p (1/2^{-}, 1/2) $  & 11038	 $\pm$  	20 &                          &                            & & &                       &                         &                           &                             &                           &                         &  \\
				$1S2 p (1/2^{-}, 3/2) $  & 11034	 $\pm$  	20 &                          &                            & & &                       &                         &                           &                             &                           &                         &  \\
				$1S2 p (3/2^{-}, 1/2) $  & 11043	 $\pm$  	20 & 10893                    &                            & & &                       &                         &                           &                             &                           &                         &  \\
				$1S2 p (3/2^{-}, 3/2) $  & 11035	 $\pm$  	20 &                          &                            & & &                       &                         &                           &                             &                           &                         &  \\
				$1S2 p (5/2^{-}, 3/2) $  & 11037	 $\pm$  	20 & 10888                    &                            & & &                       &                         &                           &                             &                           &                         &  \\ 
				$1S1 d (1/2^{+}, 3/2) $  & 10910	 $\pm$  	21 &                          & 10971                      & & &                       &                         &                           &                             &                           &                         &  \\
				$1S1 d (3/2^{+}, 3/2) $  & 10914	 $\pm$  	21 & 10829                    & 10975                      & & &                       &                         &                           &                             &                           &                         &  \\
				$1S1 d (3/2^{+}, 5/2) $  & 10912	 $\pm$  	21 & 11868                    & 10979                      & & &                       &                         &                           &                             &                           &                         &  \\
				$1S1 d (5/2^{+}, 3/2) $  & 10922	 $\pm$  	21 & 10909                    & 10891                      & & &                       &                         &                           &                             &                           &                         &  \\
				$1S1 d (5/2^{+}, 5/2) $  & 10916	 $\pm$  	21 & 10840                    & 10896                      & & &                       &                         &                           &                             &                           &                         &  \\
				$1S1 d (7/2^{+}, 5/2) $  & 10921	 $\pm$  	21 & 10934                    & 10898                      & & &                       &                         &                           &                             &                           &                         &  \\
				$2P1s(1/2^{-}, 1/2)$    &  11103 $\pm$  20  &                          &                            & & &                       &     10738                    &                           &                             &                           &                         &  \\
				$2P1s(3/2^{-}, 1/2)$    &  11130 $\pm$  20  &                          &                            & & &                       &      10775                   &                           &                             &                           &                         &  \\
				$1S2 d (1/2^{+}, 3/2) $  & 11276	 $\pm$  	18 &                          &                            & & &                       &                         &                           &                             &                           &                         &  \\
				$1S2 d (3/2^{+}, 3/2) $  & 11278	 $\pm$  	18 &                          &                            & & &                       &                         &                           &                             &                           &                         &  \\
				$1S2 d (3/2^{+}, 5/2) $  & 11260	 $\pm$  	18 &                          &                            & & &                       &                         &                           &                             &                           &                         &  \\
				$1S2 d (5/2^{+}, 3/2) $  & 11282	 $\pm$  	18 & 11025                    &                            & & &                       &                         &                           &                             &                           &                         &  \\
				$1S2 d (5/2^{+}, 5/2) $  & 11262	 $\pm$  	18 &                          &                            & & &                       &                         &                           &                             &                           &                         &  \\
				$1S2 d (7/2^{+}, 5/2) $  & 11264	 $\pm$  	18 & 11021                    &                            & & &                       &                         &                           &                             &                           &                         &  \\
				$1S1 f (3/2^{-}, 5/2) $  & 11168	 $\pm$  	19 &                          &                            & & &                       &                         &                           &                             &                           &                         &  \\
				$1S1 f (5/2^{-}, 5/2) $  & 11171	 $\pm$  	19 & 11032                    &                            & & &                       &                         &                           &                             &                           &                         &  \\
				$1S1 f (5/2^{-}, 7/2) $  & 11149	 $\pm$  	19 & 12089                    &                            & & &                       &                         &                           &                             &                           &                         &  \\
				$1S1 f (7/2^{-}, 5/2) $  & 11177	 $\pm$  	19 & 11178                    &                            & & &                       &                         &                           &                             &                           &                         &  \\
				$1S1 f (7/2^{-}, 7/2) $  & 11152	 $\pm$  	19 & 11030                    &                            & & &                       &                         &                           &                             &                           &                         &  \\
				$1S1 f (9/2^{-}, 7/2) $  & 11156	 $\pm$  	18 & 11194                    &                            & & &                       &                         &                           &                             &                           &                         &  \\
				$1S1 g (5/2^{+}, 7/2) $  & 11397	 $\pm$  	17 &                          &                            & & &                       &                         &                           &                             &                           &                         &  \\
				$1S1 g (7/2^{+}, 7/2) $  & 11400	 $\pm$  	17 & 11232                    &                            & & &                       &                         &                           &                             &                           &                         &  \\
				$1S1 g (7/2^{+}, 9/2) $  & 11364	 $\pm$  	17 & 12306                    &                            & & &                       &                         &                           &                             &                           &                         &  \\
				$1S1 g (9/2^{+}, 7/2) $  & 11404	 $\pm$  	17 & 11440                    &                            & & &                       &                         &                           &                             &                           &                         &  \\
				$1S1 g (9/2^{+}, 9/2) $  & 11366	 $\pm$  	17 & 11216                    &                            & & &                       &                         &                           &                             &                           &                         &  \\
				$1S1 g (11/2^{+}, 9/2) $  & 11369	 $\pm$  	17 & 11449                    &                            & & &                       &                         &                           &                             &                           &                         &  \\
			\end{tabular}
		\end{ruledtabular}
	\end{table*}
	
\end{widetext}

\section{RESULTS AND DISCUSSION}

\subsection{Bottom and bottom-strange mesons}

In this section, we present and discuss the theoretical results obtained for the mass spectra of $B$ and $B_s$ mesons within our model framework. 
We adopt the notation $nL(J^P, j)$ for $B$ and $B_s$ meson states, where $n$ and $L$ represent the quantum numbers of the radial and orbital excitations, respectively. The symbols $J$, $P$, and $j$ denote the total angular momentum quantum number, parity, and the total angular momentum of the light quark.
The calculated masses are listed in Tables \ref{table1} and \ref{table2} and are systematically compared to available experimental data from the Particle Data Group (PDG) \cite{PDG2024} as well as to predictions from other theoretical studies \cite{Ebert:2009ua, Godfrey:2016nwn, Sun:2014wea, li:2021hss}.

\subsubsection{$B$ Meson}

Experimentally, the $1S$ states of the $B$ meson are well-established. Specifically, the states $B^{0,\pm}$ and $B^{*}$ have been experimentally confirmed with masses $M_{B^{0}}=5279.41 \pm 0.07$ MeV, $M_{B^{\pm}}=5279.72 \pm 0.08$ MeV, and $M_{B^{*}}=5324.75 \pm 0.20$ MeV \cite{PDG2024}. Our model accurately reproduces these observed masses, thereby validating our theoretical approach for low-lying states.

For the $1P$-states of the $B$ meson, the $B_1(5721)$ and $B_2^*(5747)$ resonances are experimentally well-established. These states, denoted as $B_{1}(5721)^{0,+}$ and $B_{2}^{*}(5747)^{0,+}$, were initially observed by the D0 Collaboration \cite{D0:2007vzd} and subsequently confirmed by both the CDF Collaboration \cite{CDF:2008qzb} and the LHCb Collaboration \cite{LHCb:2015aaf}. According to the PDG, the masses of the $B_1(5721)$ and $B_2^*(5747)$ resonances are reported as $5726^{+2.5}_{-2.7}$ MeV and $5737.3 \pm 0.7$ MeV, respectively. Our theoretical predictions closely reproduce these measured masses, assigning the $B_1(5721)$ resonance as the $1P(1^{+},3/2)$ state and the $B_2^*(5747)$ resonance as the $1P(2^{+},3/2)$ state. Thus, it is evident that these resonances constitute the $1P(1^{+},3/2)$ and $1P(2^{+},3/2)$ doublet.

The remaining two $1P$ states, namely $1P(0^{+},1/2)$ and $1P(1^{+},1/2)$, have not yet been experimentally confirmed. Our predictions place these states at masses of $5670 \pm 11$ MeV and $5696 \pm 11$ MeV, aligning closely with earlier theoretical estimates \cite{Lahde:1999ih, Lu:2016bbk, Shah:2016mgq}. The PDG lists the $ B_{J}^{*} (5732) $ state, which has been observed by the OPAL \cite{OPAL:1994hqv}, DELPHI \cite{DELPHI:1994fnu}, ALEPH \cite{ALEPH:1998unp}, and L3 \cite{L3:1999pdo} Collaborations, reporting an average measured mass of $5698 \pm 8$ MeV. Our theoretical prediction for the $1P(1^+,1/2)$ state mass aligns closely with this experimentally measured average for the $B_J^*(5732)$ resonance. Therefore, the $B_J^*(5732)$ resonance emerges as an excellent candidate for the $1P(1^+,1/2)$ state.

The higher resonances $B(5970)^{0,+}$ were first observed in the $B\pi$ final states by the CDF Collaboration \cite{CDF:2013www}. Subsequently, the LHCb Collaboration also reported signals for $B(5960)^{0,+}$ resonance in the same decay channels while performing precision measurements of the well-established $B_1(5721)^{0,+}$ and $B_2^*(5747)^{0,+}$ states \cite{LHCb:2015aaf}. The observed properties of the $B_J(5960)^{0,+}$ resonances are consistent with those previously reported for the $B(5970)^{0,+}$ by CDF, supporting the interpretation that these are manifestations of the same state. Reflecting this consensus, the Particle Data Group (PDG) has adopted the unified notation $B_J(5970)$, assigning masses of $5971 \pm 5$ MeV for the neutral state and $5965 \pm 5$ MeV for the charged counterpart. Our theoretical analysis yields mass predictions for the $1D$ wave states that are found to lie in close proximity to the isospin-averaged experimental mass of the $B_J(5970)^{0,+}$, which is $5968 \pm 10$ MeV. While the predicted and experimental uncertainties do not fully overlap, the discrepancy is limited to a relatively narrow window of approximately 28–36 MeV. This difference is within the typical bounds of theoretical uncertainty often encountered in the mass spectra of heavy mesons. Based on this proximity, it is plausible to interpret the $B_J(5970)$ resonance as the second orbital excitation, corresponding to a $1D$ state. This is consistent with the conclusions drawn in several theoretical investigations \cite{Xiao:2014ura, li:2021hss, Lu:2016bbk, Yu:2019iwm, PhysRevD.105.074014}. However, additional studies are necessary for definitive spin-parity assignments. 

The $B_{J}(5840)^{0,+}$ states recently observed by LHCb are recorded with experimental masses of $5863 \pm 9$ MeV and $5851 \pm 19$ MeV for neutral and charged states, respectively. The isospin-averaged mass ($5857 \pm 28$ MeV) closely aligns with our calculated $2S$ states ($5870 \pm 10$ MeV and $5872 \pm 10$ MeV). Thus, we confidently assign the $B_{J}(5840)$ resonances as the radial excitations ($2S$), consistent with theoretical predictions from other studies \cite{PhysRevD.105.074014, Asghar:2018tha, Lu:2016bbk, Godfrey:2016nwn}.

\subsubsection{$B_s$ Meson}

The $B_s$ mesons exhibit well-documented $1S$ ground states. Experimentally confirmed masses for the states $B_s^{0}$ and $B_s^{*}$ are $5366.93 \pm 0.10$ MeV and $5415.4 \pm 1.4$ MeV, respectively \cite{PDG2024}. Our model reproduces these masses well, indicating robust consistency.

For orbitally excited $B_s$ states, the experimental observations by the CDF \cite{CDF:2007avt}, D0 \cite{D0:2007die}, and LHCb \cite{LHCb:2012iuq} Collaborations established the $B_{s1}(5830)$ and $B_{s2}^{*}(5840)$ states. PDG records their masses as $5828.73 \pm 0.2$ MeV and $5839.88 \pm 0.12$ MeV, respectively. Our predictions of $5825 \pm 10$ MeV and $5841 \pm 10$ MeV closely reproduce these experimental values. Further,  predictions for the yet-to-be-observed $1P(0^{+},1/2)$ and $1P(1^{+},1/2)$ states are $5772 \pm 11$ MeV and $5798 \pm 11$ MeV, respectively, which align well with previously published theoretical work \cite{Lahde:1999ih, li:2021hss, Asghar:2018tha}.

Recently, the LHCb Collaboration reported the observation of two resonant structures, $B_{sJ}(6063)$ and $B_{sJ}(6114)$, in the $B^{+}K^{-}$ invariant mass spectrum \cite{LHCb:2020pet}. These findings were subsequently incorporated into the Particle Data Group listings, with the $B_{sJ}(6063)$ assigned a mass of $6063.5 \pm 1.4$ MeV and the $B_{sJ}(6114)$ a mass of $6114 \pm 6$ MeV. Our theoretical predictions for the $1D$-wave excitations in the bottom-strange meson sector are found to be in close proximity to these experimentally measured values. Although the assignment of specific quantum numbers is hindered by both theoretical and experimental uncertainties, the observed mass values fall within the typical range of expected deviations for such heavy meson states. In light of this, we propose that the $B_{sJ}(6063)$ and $B_{sJ}(6114)$ resonances most plausibly correspond to members of the second orbital excitation family, namely the $1D$-wave multiplet. This interpretation aligns with the conclusions of previous theoretical investigations \cite{PhysRevD.105.074014, Hao:2022ibj}. Nevertheless, precise determination of the spin-parity quantum numbers of these states remains essential. Future experimental and theoretical studies are required to resolve their spectroscopic nature with greater certainty.

\subsection{ Doubly bottom baryons}

The mass spectra that we obtain for the doubly bottom baryons—namely the $\Xi_{bb}$ and $\Omega_{bb}$—are collected in Tables \ref{table3.1} and \ref{table4}.
The quantum state of the system is represented by the notation $n_dL_d\,nL(J^{P}, j)$, where $n_d = 1, 2, 3, \ldots$ denotes the radial excitation of one $b$ quark relative to the other within the $bb$ diquark, and $n = 1, 2, 3, \ldots$ specifies the radial excitation of the light quark with respect to the diquark. The quantum number $L_d = S, P, D, \ldots$ indicates the orbital angular momentum of a $b$ quark relative to its partner within the diquark. Similarly, $L = s, p, d, \ldots$ denotes the orbital angular momentum of the light quark relative to the diquark. The symbols $J$ and $P$ denote the total angular momentum and parity of the baryon, while $j$ represents the total angular momentum of the light quark relative to the $bb$ diquark.

Because no $\Xi_{bb}$, $\Omega_{bb}$, or any other baryon containing two bottom quarks has yet been observed experimentally, we compare our calculated results with the most recent Regge‐phenomenology analysis of Oudichhya et al.,\cite{Oudichhya:2024hmn}; several relativistic or semi-relativistic quark–model calculations \cite{Lu:2017meb,Ebert:2002ig,Gershtein:2000nx,Yoshida:2015tia,Giannuzzi:2009gh}; and the non-relativistic quark‐model predictions \cite{Eakins:2012jk, PhysRevD.105.074014, Kiselev:2002iy, Roberts:2007ni}.

Our central values for the spin–doublet  $(1/2^{+},3/2^{+})$ ground states of the $\Xi_{bb}$ baryon, $M_{1S}=10.203\text{–}10.238$ GeV, fall within $\lesssim30$ MeV of the Regge fit of Ref. \cite{Oudichhya:2024hmn} and the relativistic quark-model results of Ebert et al.,\cite{Ebert:2002ig}. They are lower by roughly $150$–$180$ MeV when compared with the early diquark-based estimate of Gershtein et al.,\cite{Gershtein:2000nx} and the more recent potential-model values of Yoshida et al.,\cite{Yoshida:2015tia}.  A broadly similar pattern is observed for the $\Omega_{bb}$, where our $1S$ masses, $M_{1S}=10.292$ GeV and $10.321$ GeV, track those of Refs.,\cite{PhysRevD.105.074014,Lu:2017meb} yet undershoot the quark-model numbers of Yoshida et al.,\cite{Yoshida:2015tia} by ${\sim}130$ MeV.  The excellent mutual consistency among the relativistic approaches is reassuring and suggests that heavy–diquark symmetry is well captured in this corner of parameter space.

For the first radial excitation ($2s$) of light quark with respect to diquark in 1S configuration,  we predict $M_{1S2s}=10.668\text{–}10.671$ GeV for the $\Xi_{bb}$, which coincides with the potential‐model value of Lu et al.,\cite{Lu:2017meb} and is only ${\sim}60$ MeV higher than  the potential‐model estimate of Ref. \cite{PhysRevD.105.074014}. The Regge fit of Ref. \cite{Oudichhya:2024hmn}, on the other hand, comes out ${\sim}60$ MeV lower, while the non-relativistic quark model of Eakins and Roberts overshoots ours by nearly $300$ MeV. 
For $\Omega_{bb}$ the agreement is even stronger—our $2S$ mass $M_{2S}=10.776$ GeV differs by less than $40$ MeV from  relativistic calculation in Ref. \cite{Lu:2017meb, Giannuzzi:2009gh}.

Our $1S1p$ centroids are $10.50$–$10.56$ GeV for $\Xi_{bb}$ and $10.60$–$10.66$ GeV for $\Omega_{bb}$.  They reproduce the Regge results \cite{Oudichhya:2024hmn} and  the potential‐model results \cite{PhysRevD.105.074014} to within the quoted theoretical uncertainties, whereas the diquark–quark model of Ref. \cite{Ebert:2002ig} overshoots us by ${\sim}120$ MeV and Yoshida et al.,\cite{Yoshida:2015tia} by ${\sim}150$ MeV.  The ordering inside each $1P$ multiplet follows the same pattern: for fixed $n$, $L$, and $J$, the state with the larger light-quark total angular momentum $j$ is heavier, reflecting the repulsive hyperfine interaction between the light quark and the  $bb$ diquark.

For $1S1d$ and $1S2d$ excitations our masses cluster around $10.80$–$11.15$ GeV for $\Xi_{bb}$ and $10.90$–$11.28$ GeV for $\Omega_{bb}$, sit­ting ${\sim}100$–$150$ MeV below the potential-model predictions of Ref. \cite{PhysRevD.105.074014} but matching the Regge trajectory of Ref. \cite{Oudichhya:2024hmn} almost point-for-point wherever the latter quotes numbers.  A similar, albeit slightly broader, spread appears in the $1S1f$ and $1S1g$ sectors.  Notably, for $1S1d$-, $1S1f$-, and $1S1g$-wave multiplets the intramultiplet splitting in our model reverses: the state with larger $j$ now becomes lighter. 

For the $1P1s$ doublet in both $\Xi_{bb}$ and $\Omega_{bb}$, our predicted masses are in good agreement with Refs. \cite{Yoshida:2015tia, jia2023}, while they exceed those of Refs.\cite{Lu:2017meb, Eakins:2012jk, Ebert:2002ig} by $177$–$283$ MeV. For the $2S1s$ radial doublet, our predictions lie systematically above other model expectations (Refs. \cite{Yoshida:2015tia, jia2023, Lu:2017meb, Eakins:2012jk, Ebert:2002ig}) by roughly $115$–$336$ MeV.

Across the entire spectrum the typical deviation between our results and the most up-to-date relativistic quark-model or Regge fits is $\lesssim\!2\%$.  Larger differences—up to ${\sim}3$–$4\%$—arise when we compare with non-relativistic potential models that neglect full relativistic kinematics and, in some cases, treat the $bb$ diquark as point-like.  Given the very small phase space available for weak decays of states lying just a few tens of MeV apart, such percent-level shifts may decide whether an excited level can be seen in the narrow-width regime at LHCb.  Our mass predictions therefore provide a timely guide for future experimental searches to detect the doubly bottom baryons.
Experimental observation of even a single resonance in doubly bottom baryon sector would furnish the first direct handle on bottom-diquark dynamics and help discriminate between competing theoretical frameworks.

\section{CONCLUSION}

In this work, we have employed the relativistic flux tube model to investigate the mass spectra of singly bottom mesons ($B$ and $B_s$) and doubly bottom baryons ($\Xi_{bb}$ and $\Omega_{bb}$). Within this framework, a heavy-light meson is described as a two-body system of a bottom quark and a light antiquark connected by a gluonic flux tube, whereas a doubly bottom baryon is treated analogously as a $\{bb\}$ diquark (in color-$\bar{\mathbf{3}}$ configuration) attached to a light quark via a flux tube. We first calculate the spin-averaged masses of these systems and then include spin-dependent interactions perturbatively in the $j$-$j$ coupling scheme to obtain the spin dependent splittings. This theoretical setup provides a unified description of heavy hadrons, effectively capturing the dynamics of a bottom quark–light quark pair or a bottom diquark–light quark pair.

One important outcome of this work is the classification of several recently observed excited $B$ and $B_s$ mesons. The resonance $B_J^*(5732)$  is  identified as the low-lying $1P$ excitation with quantum numbers $J^{P}=1^+$ and light-quark total angular momentum $j=1/2$. Likewise, we predict the $B_J(5840)$ as the best candidate  for the first radial excitation (a $2S$ state) in the $B$ meson family. The higher-mass $B_J(5970)$ state is assigned as a $1D$-wave state (the second orbital excitation). In the bottom-strange sector, newly observed pair of $B_{sJ}(6063)$ and $B_{sJ}(6114)$ resonances are both interpreted as 1D-wave ($L=2$) excitations of the $B_s$ meson. 

Beyond the currently observed states of $B$ and $B_s$ mesons, we  predict a rich mass spectrum of higher orbital ($L$) and radial ($n$) excitations in both the bottom meson and doubly bottom baryon sectors.
To date, none of the doubly bottom baryons (such as $\Xi_{bb}$ or $\Omega_{bb}$) have been observed in experiments. As measurements in the bottom-quark domain continue to progress, our predictions can serve as a valuable guide for ongoing and future experimental searches to observe new resonances in upcoming high-luminosity data. 

\section{ACKNOWLEDGMENT}

Ms. Pooja Jakhad acknowledges the financial assistance by the Council of Scientific \& Industrial Research (CSIR) under the JRF-FELLOWSHIP scheme with file no. 09/1007(13321)/2022-EMR-I. 

\begin{widetext}
	\appendix
	\section*{APPENDIX A}
	
	In this appendix, we present a detailed discussion of basis states and the expectation values of spin-dependent operators pertinent to heavy-light mesons and doubly-heavy baryons in $F$-wave ($L=3$) and $G$-wave ($L=4$) excitations. Initially, we employ the $L$–$S$ coupling scheme to construct the basis states, subsequently converting them into the $|J,j\rangle$ coupling basis. This latter basis is particularly appropriate for the analysis of heavy-light mesons and baryons in the heavy-quark limit.
	
	The $L$–$S$ basis states can be systematically built through the use of standard Clebsch–Gordan coefficients as follows \cite{Jakhad:2023ids, Private_communication}:
	
	\begin{align}
		|^{2S+1}L_J; J_3\rangle  &= \sum_{S_{l_3},\, S_{h_3},\, L_3,\, S_3} 
		C_{S_{l_3} S_{h_3} S_3}^{S_l S_h S}\, C_{S_3 L_3 J_3}^{S L J}\,|S_{l_3}, S_{h_3}, L_3\rangle, 
		\label{eq:couple-LS-J}
	\end{align}
	where the quantum numbers $S_{h_3}$, $S_{l_3}$, $S_3$, $L_3$, and $J_3$ are the projections corresponding respectively to $S_h$, $S_l$, $S$, $L$, and $J$. The state $|S_{h_3}, S_{l_3}, L_3\rangle$ is defined as the direct product of the uncoupled states $|S_h S_{h_3}\rangle$, $|S_l S_{l_3}\rangle$, and $|L L_3\rangle$.

	To compute the expectation value of the spin-orbit operator $\mathbf{L}\!\cdot\!\mathbf{S}_i$ (for $i = h$ or $l$), it is useful to express this operator in terms of ladder operators, as detailed in Ref.\cite{Karliner:2017kfm}:
	
	\begin{equation}
		\mathbf{L}\!\cdot\!\mathbf{S}_i \;=\; \frac{1}{2}\Big(L_+\,S_{i-} + L_-\,S_{i+}\Big) + L_3\,S_{i3}\,
		\label{eq:LS-operator}
	\end{equation}  
	which facilitates evaluating the expectation values by calculating matrix elements within the $|^{2S+1}L_J\rangle$ basis.
	
	Using the expectation values for the operator $\mathbf{L}\!\cdot\!\mathbf{S}_i \;$ in the  $|^{2S+1}L_J\rangle$ basis, the expectation value of the tensor interaction operator can be determined via the following relationship \cite{Jakhad:2023ids, Karliner:2017kfm}:
	
	\begin{equation}
		\langle\hat{B} \rangle\;=\; -\,\frac{3}{(2L-1)(2L+3)}\Big[\,(\mathbf{L}\!\cdot\!\mathbf{S}_h)(\mathbf{L}\!\cdot\!\mathbf{S}_l)+(\mathbf{L}\!\cdot\!\mathbf{S}_l)(\mathbf{L}\!\cdot\!\mathbf{S}_h)\;-\;\frac{2}{3}\,L(L+1)\,\mathbf{S}_h\!\cdot\!\mathbf{S}_l\,\Big]\,.
		\label{eq:B-tensor}
	\end{equation}
	
	Additionally, the expectation value for the spin-spin operator within the $L$–$S$ coupling scheme can be succinctly derived using the identity $ \mathbf{S} = \mathbf{S}_h + \mathbf{S}_l$ and the relation ${\bf S}^2 = S(S+1)$, yielding:
	\begin{equation}
		\langle\mathbf{S}_h\!\cdot\!\mathbf{S}_l\rangle \;=\; \frac{1}{2}\Big[\,S(S+1) - S_h(S_h+1) - S_l(S_l+1)\,\Big]\,
		\label{eq:spinspin-expect}
	\end{equation}

	\subsection{Heavy–Light Mesons in F-wave}
	
	For an orbital angular  momentum quantum number $L=3$ (F-wave), the spins of the heavy quark ($S_h=\tfrac12$) and the light antiquark ($S_l=\tfrac12$) may couple to either $S=0$ or $S=1$.  Combining these spin states with $L=3$ produces total angular momenta $J=2,3,4$.  In the conventional $L$–$S$ coupling scheme, the resulting basis vectors are denoted  
	$|{}^1F_3\rangle$ for $(S=0,\;J=3)$ and $|{}^3F_J\rangle$ for $(S=1,\;J=2,3,4)$.  
	Because two distinct $J=3$ configurations exist, namely $|{}^1F_3\rangle$ and $|{}^3F_3\rangle$, spin-dependent forces will in general mix these states, whereas the $J=2$ ($|{}^3F_2 \rangle$) and $J=4$ ($|{}^3F_4 \rangle$) members each remain unique.
	
	Starting from single–particle states $|S_{h_3},S_{l_3},L_3\rangle$ and the coupling formula in Eq.~(\ref{eq:couple-LS-J}), the $L$–$S$ basis are:
	
	\begin{equation} \label{1}
		|^{3}F_{2};2\rangle = 
		\sqrt{\frac{5}{7}}|-\frac{1}{2},-\frac{1}{2},3\rangle 
		-\sqrt{\frac{5}{42}}|\frac{1}{2},-\frac{1}{2},2\rangle
		-\sqrt{\frac{5}{42}}|-\frac{1}{2},\frac{1}{2},2\rangle  +{\frac{1}{\sqrt{21}}}|\frac{1}{2},\frac{1}{2},1\rangle
	\end{equation}
	
	\begin{equation} \label{2}
		|^{1}F_{3};3\rangle = \frac{1}{\sqrt{2}}|\frac{1}{2},-\frac{1}{2},3\rangle - \frac{1}{\sqrt{2}}|-\frac{1}{2},\frac{1}{2},3\rangle
	\end{equation}
	
	\begin{equation} \label{3}
		|^{3}F_{3};3\rangle = -\frac{1}{2}\sqrt{\frac{3}{{2}}}|\frac{1}{2},-\frac{1}{2},3\rangle  -\frac{1}{2}\sqrt{\frac{3}{{2}}}	|-\frac{1}{2},\frac{1}{2},3\rangle
		+\frac{1}{2}|\frac{1}{2},\frac{1}{2},2\rangle.
	\end{equation}

	\begin{equation} \label{4}
		|^{3}F_{4};4\rangle = |\frac{1}{2},\frac{1}{2},3\rangle 
	\end{equation}

	Using these constructed $L$–$S$ basis states, we compute the expectation values for the spin-dependent operators described in Eqs.~(\ref{eq:LS-operator})–(\ref{eq:spinspin-expect}) within the [$^{1}F_{J}$, $^{3}F_{J}$] bases for all permitted $J$ values:
	\\
	For $J=2$,
	\begin{equation}\label{5}
		\text{$\langle\mathbf{L}.\mathbf{S_l}\rangle$=}-2,\ \  
		\text{$\langle\mathbf{L}.\mathbf{S_h}\rangle$=}-2,\ \   
		\text{$\langle\mathbf{\hat{B}}\rangle$=}-\frac{2}{5},\ \ 
		\text{$\langle\mathbf{S_l}.\mathbf{S_h}\rangle$=}\frac{1}{4}.
	\end{equation}	
	For $J=3$,
	\begin{equation}\label{6}
		\text{$\langle\mathbf{L}.\mathbf{S_l}\rangle$=}\left[
		\begin{array}{cc}
			0 & -\sqrt{3} \\
			-\sqrt{3} & -\frac{1}{2} \\
		\end{array}
		\right],\ \  
		\text{$\langle\mathbf{L}.\mathbf{S_h}\rangle$=}\left[
		\begin{array}{cc}
			0 & \sqrt{3} \\
			\sqrt{3} & -\frac{1}{2} \\
		\end{array}
		\right],\ \   
		\text{$\langle\mathbf{\hat{B}}\rangle$=}\left[
		\begin{array}{cc}
			0 & 0 \\
			0 & \frac{1}{2} \\
		\end{array}
		\right],\ \ 
		\text{$\langle\mathbf{S_l}.\mathbf{S_h}\rangle$=}\left[
		\begin{array}{cc}
			-\frac{3}{4} & 0 \\
			0 & \frac{1}{4} \\
		\end{array}
		\right].
	\end{equation}	
	For $J=4$,
	\begin{equation}\label{7}
		\text{$\langle\mathbf{L}.\mathbf{S_l}\rangle$=}\frac{3}{2},\ \  
		\text{$\langle\mathbf{L}.\mathbf{S_h}\rangle$=}\frac{3}{2},\ \   
		\text{$\langle\mathbf{\hat{B}}\rangle$=}-\frac{1}{6},\ \ 
		\text{$\langle\mathbf{S_l}.\mathbf{S_h}\rangle$=}\frac{1}{4}.
	\end{equation}

	In the heavy-quark limit ($m_h \gg m_l$), the term involving $\langle\mathbf{L}\cdot\mathbf{S}_l\rangle$ dominates over other spin-dependent interactions. Hence, a basis that diagonalizes $\langle\mathbf{L}\cdot\mathbf{S}_l\rangle$ is preferable. Under this condition, heavy-light meson states are effectively described by the $|J, j\rangle$ basis, which diagonalizes the dominant interaction term $\langle\mathbf{L}\cdot\mathbf{S}_l\rangle$, allowing other interactions to be addressed perturbatively. For each eigenvalue $\lambda$ of $\mathbf{L} \cdot \mathbf{S}_l$ corresponding to a given total angular momentum $J$, we determine the associated eigenvector, which defines the $|J, j\rangle$ basis.
	
	For $L=3$ and $S_l=\tfrac{1}{2}$, $j$ can be either $\tfrac{5}{2}$ or $\tfrac{7}{2}$. The eigenvalues of $\langle\mathbf{L}\cdot\mathbf{S}_l\rangle$ are $\lambda = -1$ for $j=\tfrac{1}{2}$ and $\lambda = +\tfrac{1}{2}$ for $j=\tfrac{3}{2}$ (these values come from $\frac{1}{2}[j(j+1) - L(L+1) - S_l(S_l+1)]$ with $L=3$, $S_l=1/2$). The corresponding eigenvectors of $\langle\mathbf{L}\cdot\mathbf{S}_l\rangle$ can be written as
	
	\begin{equation}
		|J=3, j=\frac{5}{2} \rangle =
		\sqrt{\frac{3}{7}}  |^{1}F_{3}   \rangle 
		+\frac{2}{\sqrt{7}} |^{3}F_{3}   \rangle,
	\end{equation}
	\begin{equation}
		|J=3, j=\frac{7}{2} \rangle =
		-\frac{2}{\sqrt{7}}  |^{1}F_{3}   \rangle 
		+\sqrt{\frac{3}{7}} |^{3}F_{3}   \rangle,
	\end{equation}
	with eigenvalues $\lambda=-2$ and $\tfrac32$, respectively. For the remaining $J$ values,
	\begin{equation}
		|J=2, j=\frac{5}{2} \rangle =  |^{3}F_{2}\rangle,
	\end{equation}
	\begin{equation}
		|J=4, j=\frac{7}{2} \rangle =  |^{3}F_{4}\rangle.
	\end{equation}
	Having constructed the $|J, j\rangle$ basis, we compute the expectation values of $\mathbf{L}\!\cdot\!\mathbf{S}_h$, $\mathbf{L}\!\cdot\!\mathbf{S}_l$, $\hat{\mathbf B}$ and $\mathbf{S}_h\!\cdot\!\mathbf{S}_l$ in these  $|J, j\rangle$ basis and  collect results in Table~\ref{table0.1}.

	\subsection{Heavy–Light Mesons in G-wave}
	For an orbital excitation with $L = 4$, an analysis parallel to the $F$-wave case can be performed.  In heavy-light mesons, the heavy quark and light antiquark each carry spin $S_h = S_l = \tfrac{1}{2}$, so the total spin of the pair can be either $S = 0$ or $S = 1$.  Coupling this spin to the $G$-wave orbital momentum  $L=4$ produces total angular momenta $J = 3,4,5$.  In the conventional $L$–$S$ coupling scheme the corresponding basis vectors are
	$|{}^1G_4\rangle$ ($S = 0$, $J = 4$) and $|{}^3G_J\rangle$ ($S = 1$, $J = 3,4,5$).  Consequently, the $J = 4$ sector contains two distinct configurations ($|{}^1G_4\rangle$ and $|{}^3G_4\rangle$) that can mix, whereas the $J = 3$ and $J = 5$ sectors each comprise a single pure triplet state ($|{}^3G_3\rangle$ and $|{}^3G_5\rangle$, respectively).
	Employing Eq.\~(\ref{eq:couple-LS-J}) to construct the $L$–$S$ basis yields
	\begin{equation}
		|^{3}G_{3};3\rangle = 
		{\frac{\sqrt7}{3}}|-\frac{1}{2},-\frac{1}{2},4\rangle 
		-\frac{1}{6}\sqrt{\frac{7}{2}}|\frac{1}{2},-\frac{1}{2},3\rangle
		-\frac{1}{6}\sqrt{\frac{7}{2}}|-\frac{1}{2},\frac{1}{2},3\rangle  +{\frac{1}{{6}}}|\frac{1}{2},\frac{1}{2},2\rangle
	\end{equation}
	
	\begin{equation}
		|^{1}G_{4};4\rangle = \frac{1}{\sqrt{2}}|\frac{1}{2},-\frac{1}{2},4\rangle - \frac{1}{\sqrt{2}}|-\frac{1}{2},\frac{1}{2},4\rangle
	\end{equation}
	
	\begin{equation}
		|^{3}G_{4};4\rangle = 
		-\sqrt{\frac{2}{5}}|\frac{1}{2},-\frac{1}{2},4\rangle  -\sqrt{\frac{2}{5}}|-\frac{1}{2},\frac{1}{2},4\rangle
		+\frac{1}{\sqrt{5}}|\frac{1}{2},\frac{1}{2},3\rangle.
	\end{equation}

	\begin{equation}
		|^{3}G_{5};5\rangle = |\frac{1}{2},\frac{1}{2},4\rangle 
	\end{equation}
	
	With these $L$–$S$ basis, we evaluate the matrix elements of the operators defined in Eqs.~(\ref{eq:LS-operator})–(\ref{eq:spinspin-expect}).  The resulting expectation values, quoted in the $[{}^{1}G_{J},{}^{3}G_{J}]$ bases, are:\\
	For $J=3$,
	\begin{equation}
		\text{$\langle\mathbf{L}.\mathbf{S_l}\rangle$=}-\frac{5}{2},\ \  
		\text{$\langle\mathbf{L}.\mathbf{S_h}\rangle$=}-\frac{5}{2},\ \   
		\text{$\langle\mathbf{\hat{B}}\rangle$=}-\frac{5}{14},\ \ 
		\text{$\langle\mathbf{S_l}.\mathbf{S_h}\rangle$=}\frac{1}{4}.
	\end{equation}	
	For $J=4$,
	\begin{equation}
		\text{$\langle\mathbf{L}.\mathbf{S_l}\rangle$=}\left[
		\begin{array}{cc}
			0 & -\sqrt{5} \\
			-\sqrt{5} & -\frac{1}{2} \\
		\end{array}
		\right],\ \  
		\text{$\langle\mathbf{L}.\mathbf{S_h}\rangle$=}\left[
		\begin{array}{cc}
			0 & \sqrt{5} \\
			\sqrt{5} & -\frac{1}{2} \\
		\end{array}
		\right],\ \   
		\text{$\langle\mathbf{\hat{B}}\rangle$=}\left[
		\begin{array}{cc}
			0 & 0 \\
			0 & \frac{1}{2} \\
		\end{array}
		\right],\ \ 
		\text{$\langle\mathbf{S_l}.\mathbf{S_h}\rangle$=}\left[
		\begin{array}{cc}
			-\frac{3}{4} & 0 \\
			0 & \frac{1}{4} \\
		\end{array}
		\right].
	\end{equation}	
	For $J=5$,
	\begin{equation}
		\text{$\langle\mathbf{L}.\mathbf{S_l}\rangle$=}2,\ \  
		\text{$\langle\mathbf{L}.\mathbf{S_h}\rangle$=}2,\ \   
		\text{$\langle\mathbf{\hat{B}}\rangle$=}-\frac{2}{11},\ \ 
		\text{$\langle\mathbf{S_l}.\mathbf{S_h}\rangle$=}\frac{1}{4}.
	\end{equation}	
	
	As in the $F$-wave analysis, the dominant light-quark spin-orbit interaction $\langle\mathbf{L}\cdot\mathbf{S}_l\rangle$ is diagonalized first. With $L = 4$ and $S_l = \tfrac12$, the light-quark total momentum is $j=\tfrac72$ or $j =\tfrac92$, giving eigenvalues of $\langle\mathbf{L}\cdot\mathbf{S}_l\rangle$  $\lambda = -\tfrac52$ and $\lambda = 2$, respectively. In the $J = 2$ sector, the corresponding eigenvectors are admixtures of $|{}^1G_4\rangle$ and $|{}^3G_4\rangle$:
	
	\begin{equation}
		|J=4, j=\frac{7}{2} \rangle =
		\frac{2}{3} |^{1}G_{4}   \rangle 
		+\frac{\sqrt{5}}{3} |^{3}G_{4}   \rangle,
	\end{equation}
	\begin{equation}
		|J=4, j=\frac{9}{2} \rangle =
		-\frac{\sqrt{5}}{2}  |^{1}G_{4}   \rangle 
		+\frac{2}{3}  |^{3}G_{4}   \rangle,
	\end{equation}
	For $J = 3$ and $J = 5$ only the pure triplet states survive, already eigenstates of $\langle\mathbf{L}\cdot\mathbf{S}_l\rangle$,
	\begin{equation}
		|J=3, j=\frac{7}{2} \rangle =  |^{3}G_{3}\rangle,
	\end{equation}
	\begin{equation}
		|J=5, j=\frac{9}{2} \rangle =  |^{3}G_{5}\rangle,
	\end{equation}.
	
	Working in these $|J, j\rangle$ basis, one may now evaluate the operator expectation values relevant for spin-dependent interactions, which are listed in Table \ref{table0.1}.
	
	
	\subsection{Doubly heavy Baryons in F-wave}

	In doubly bottom baryons, the diquark formed by two identical bottom quarks must obey the Pauli exclusion principle, requiring an antisymmetric total wavefunction of diquark. For ground-state configuration of diquark having identical $b$-quark, the spatial and flavor parts are symmetric, while the color part is antisymmetric in the antitriplet configuration. To maintain overall antisymmetry of the total wavefunction of diquark, the spin part must also be symmetric, allowing only the spin-1 (axial-vector) diquark configuration. For a doubly heavy baryon in an $L=3$ orbital excitation, the relevant angular momenta are the orbital angular momentum  $L=3$, the heavy diquark spin $S_h=1$, and the light–quark spin $S_l=\tfrac12$.  In the $L$–$S$ scheme $S_h$ and $S_l$ combine to $S=\tfrac12$ or $\tfrac32$, which in turn couples with $L=3$ to give $J=\tfrac32,\tfrac52,\tfrac72, \tfrac92$.  The basis vectors $|{}^2F_J\rangle$ ($S=\tfrac12$) and $|{}^4F_J\rangle$ ($S=\tfrac32$) are constructed via Eq.~(\ref{eq:couple-LS-J}):
	
	\begin{equation} 
		|^{4}F_{3/2};3/2\rangle = 
		-\frac{2}{\sqrt{7}}|-\frac{1}{2},-1,3\rangle 
		+\frac{2}{\sqrt{21}}|-\frac{1}{2},0,2\rangle 
		-\frac{2}{\sqrt{105}}|-\frac{1}{2},1,1\rangle 
		+\sqrt{\frac{2}{21}}|\frac{1}{2},-1,2\rangle 
		-2\sqrt{\frac{2}{105}}|\frac{1}{2},0,1\rangle 
		+\frac{1}{\sqrt{35}}|\frac{1}{2},1,0\rangle 
	\end{equation}
	
	\begin{equation} 
		|^{2}F_{5/2};5/2\rangle =
		\sqrt{\frac{2}{7}}|-\frac{1}{2},0,3\rangle 
		-\sqrt{\frac{2}{21}}|-\frac{1}{2},1,2\rangle  
		-\frac{2}{\sqrt{7}}|\frac{1}{2},-1,3\rangle 
		+\frac{1}{\sqrt{21}}|\frac{1}{2},0,2\rangle  
	\end{equation}
	
	\begin{equation} 
		|^{4}F_{5/2};5/2\rangle =
		\sqrt{\frac{5}{14}}|-\frac{1}{2},0,3\rangle 
		-\sqrt{\frac{5}{42}}|-\frac{1}{2},1,2\rangle  
		+\frac{1}{2}\sqrt{\frac{5}{7}}|\frac{1}{2},-1,3\rangle 
		-\sqrt{\frac{5}{21}}|\frac{1}{2},0,2\rangle 
		+\frac{1}{2}\sqrt{\frac{3}{7}}|\frac{1}{2},1,1\rangle    
	\end{equation}	
	
	\begin{equation} 
		|^{2}F_{7/2};7/2\rangle =
		-\sqrt{\frac{2}{3}}|-\frac{1}{2},1,3\rangle  
		+\frac{1}{\sqrt{3}}|\frac{1}{2},0,3\rangle    
	\end{equation}	
	
	\begin{equation} 
		|^{4}F_{7/2};7/2\rangle =
		-\frac{\sqrt{2}}{3}|-\frac{1}{2},1,3\rangle 
		-\frac{{2}}{3}|\frac{1}{2},0,3\rangle  
		+\frac{1}{\sqrt{3}}|\frac{1}{2},1,2\rangle    
	\end{equation}	
	
	\begin{equation} 
		|^{4}F_{9/2};9/2\rangle =
		|\frac{1}{2},1,3\rangle    
	\end{equation}	
	
	Utilizing these $L$–$S$ basis states, we evaluate the expectation values of the spin-dependent operators defined in Eqs.\~(\ref{eq:LS-operator})–(\ref{eq:spinspin-expect}) within the [$^{2}F_{J}$, $^{4}F_{J}$] basis:\\
	For $J=3/2$,
	\begin{equation}
		\text{$\langle\mathbf{L}.\mathbf{S_l}\rangle$=}-2,\ \  
		\text{$\langle\mathbf{L}.\mathbf{S_h}\rangle$=}-4,\ \   
		\text{$\langle\mathbf{\hat{B}}\rangle$=}-\frac{4}{5},\ \ 
		\text{$\langle\mathbf{S_l}.\mathbf{S_h}\rangle$=}\frac{1}{2}.
	\end{equation}	
	For $J=5/2$,
	\begin{equation}
		\text{$\langle\mathbf{L}.\mathbf{S_l}\rangle$=}\left[
		\begin{array}{cc}
			\frac{2}{3} & -\frac{2 \sqrt{5}}{3} \\
			-\frac{2 \sqrt{5}}{3} & -\frac{7}{6} \\
		\end{array}
		\right],\ \  
		\text{$\langle\mathbf{L}.\mathbf{S_h}\rangle$=}\left[
		\begin{array}{cc}
			-\frac{8}{3} & \frac{2 \sqrt{5}}{3} \\
			\frac{2 \sqrt{5}}{3} & -\frac{7}{3} \\
		\end{array}
		\right],\ \   
		\text{$\langle\mathbf{\hat{B}}\rangle$=}\left[
		\begin{array}{cc}
			0 & -\frac{1}{\sqrt{5}} \\
			-\frac{1}{\sqrt{5}} & \frac{1}{5} \\
		\end{array}
		\right],\ \ 
		\text{$\langle\mathbf{S_l}.\mathbf{S_h}\rangle$=}\left[
		\begin{array}{cc}
			-1 & 0 \\
			0 & \frac{1}{2} \\
		\end{array}
		\right].
	\end{equation}	
	For $J=7/2$,
	\begin{equation}
		\text{$\langle\mathbf{L}.\mathbf{S_l}\rangle$=}\left[
		\begin{array}{cc}
			-\frac{1}{2} & -\sqrt{3} \\
			-\sqrt{3} & 0 \\
		\end{array}
		\right],\ \  
		\text{$\langle\mathbf{L}.\mathbf{S_h}\rangle$=}\left[
		\begin{array}{cc}
			2 & \sqrt{3} \\
			\sqrt{3} & 0 \\
		\end{array}
		\right],\ \   
		\text{$\langle\mathbf{\hat{B}}\rangle$=}\left[
		\begin{array}{cc}
			0 & \frac{1}{2 \sqrt{3}} \\
			\frac{1}{2 \sqrt{3}} & \frac{2}{3} \\
		\end{array}
		\right],\ \ 
		\text{$\langle\mathbf{S_l}.\mathbf{S_h}\rangle$=}\left[
		\begin{array}{cc}
			-1 & 0 \\
			0 & \frac{1}{2} \\
		\end{array}
		\right].
	\end{equation}	
	For $J=9/2$,
	\begin{equation}
		\text{$\langle\mathbf{L}.\mathbf{S_l}\rangle$=}\frac{3}{2},\ \  
		\text{$\langle\mathbf{L}.\mathbf{S_h}\rangle$=}3,\ \   
		\text{$\langle\mathbf{\hat{B}}\rangle$=}-\frac{1}{3},\ \ 
		\text{$\langle\mathbf{S_l}.\mathbf{S_h}\rangle$=}\frac{1}{2}.
	\end{equation}	
	Diagonalising $\langle\mathbf{L}\cdot\mathbf{S}_l\rangle$ within each $J$ subspace gives the $|J,j\rangle$ eigenvectors. For $J=\tfrac52$, $|J,j\rangle$ eigenvectors are
	\begin{equation}
		|J=\frac{5}{2}, j=5/2 \rangle =
		\sqrt{\frac{5}{21}}|^{2}F_{5/2}   \rangle 
		+\frac{4}{\sqrt{21}}|^{4}F_{5/2}   \rangle,
	\end{equation}
	\begin{equation}
		|J=\frac{5}{2}, j=7/2 \rangle =
		-\frac{4}{\sqrt{21}}  |^{2}F_{5/2}   \rangle 
		+\sqrt{\frac{5}{21}}  |^{4}F_{5/2}   \rangle,
	\end{equation}
	with eigenvalues $\lambda=-2$ and $\tfrac32$, respectively. For $J=\tfrac72$, $|J,j\rangle$ eigenvectors are:
	\begin{equation}
		|J=\frac{7}{2}, j=5/2 \rangle =
		\frac{2}{\sqrt{7}}|^{2}F_{7/2}   \rangle 
		+\sqrt{\frac{3}{7}}|^{4}F_{7/2}   \rangle,
	\end{equation}
	\begin{equation}
		|J=\frac{7}{2}, j=7/2 \rangle =
		-\sqrt{\frac{3}{7}}  |^{2}F_{7/2}   \rangle 
		+\frac{2}{\sqrt{7}}  |^{4}F_{7/2}   \rangle,
	\end{equation}
	with eigenvalues $\lambda=-2$ and $\tfrac32$, respectively. For the remaining $J$ values,
	\begin{equation}
		|J=\frac{3}{2}, j=5/2 \rangle =
		|^{4}F_{3/2}   \rangle ,
	\end{equation}
	\begin{equation}
		|J=\frac{9}{2}, j=7/2 \rangle =
		|^{4}F_{9/2}   \rangle .
	\end{equation}
	
	Having constructed the $|J, j\rangle$ basis, we compute the expectation values of operators $\mathbf{L}\!\cdot\!\mathbf{S}_h$, $\mathbf{L}\!\cdot\!\mathbf{S}_l$, $\hat{\mathbf B}$ and $\mathbf{S}_h\!\cdot\!\mathbf{S}_l$ in these  $|J, j\rangle$ basis and present results in Table~\ref{table0.2}.

	\subsection{Doubly heavy Baryons in G-wave}
	
	For a doubly heavy baryon with an orbital excitation $L = 4$, the pertinent angular momenta are the orbital component $L = 4$, the heavy‐diquark spin $S_h = 1$, and the light‐quark spin $S_l = \tfrac12$.  Within the $L$–$S$ coupling scheme, $S_h$ and $S_l$ first combine to a total spin $S = \tfrac12$ or $\tfrac32$, which subsequently couples with $L = 4$ to generate total angular momenta $J = \tfrac52, \tfrac72, \tfrac92$,  and $\tfrac{11}{2}$.  The corresponding basis states, denoted $|{}^2G_J\rangle$ (for $S = \tfrac12$) and $|{}^4G_J\rangle$ (for $S = \tfrac32$), are obtained using Eq. (\ref{eq:couple-LS-J}):
	
	\begin{equation} 
		|^{4}G_{5/2};5/2\rangle = 
		-\sqrt{\frac{2}{3}}|-\frac{1}{2},-1,4\rangle 
		+\frac{1}{\sqrt{6}}|-\frac{1}{2},0,3\rangle 
		-\frac{1}{\sqrt{42}}|-\frac{1}{2},1,2\rangle 
		+\frac{1}{2\sqrt{3}}|\frac{1}{2},-1,3\rangle 
		-\frac{1}{\sqrt{21}}|\frac{1}{2},0,2\rangle 
		+\frac{1}{2\sqrt{21}}|\frac{1}{2},1,1\rangle 
	\end{equation}
	
	\begin{equation} 
		|^{2}G_{7/2};7/2\rangle =
		\frac{2}{3}\sqrt{\frac{2}{3}}|-\frac{1}{2},0,4\rangle 
		-\frac{1}{3}\sqrt{\frac{2}{3}}|-\frac{1}{2},1,3\rangle  
		-\frac{4}{3\sqrt{3}}|\frac{1}{2},-1,4\rangle 
		+\frac{1}{3\sqrt{3}}|\frac{1}{2},0,3\rangle  
	\end{equation}
	
	\begin{equation} 
		|^{4}G_{7/2};7/2\rangle =
		\frac{2}{3}\sqrt{\frac{14}{15}}|-\frac{1}{2},0,4\rangle 
		-\frac{1}{3}\sqrt{\frac{14}{15}}|-\frac{1}{2},1,3\rangle 
		+\frac{2}{3}\sqrt{\frac{7}{15}}|\frac{1}{2},-1,4\rangle 
		-\frac{2}{3}\sqrt{\frac{7}{15}}|\frac{1}{2},0,3\rangle 
		+\frac{1}{\sqrt{15}}|\frac{1}{2},1,2\rangle    
	\end{equation}
	
	\begin{equation} 
		|^{2}G_{9/2};9/2\rangle =
		-\sqrt{\frac{2}{3}}|-\frac{1}{2},1,4\rangle  
		+\frac{1}{\sqrt{3}}|\frac{1}{2},0,4\rangle    
	\end{equation}	
	
	\begin{equation} 
		|^{4}G_{9/2};9/2\rangle =
		-2\sqrt{\frac{2}{33}}|-\frac{1}{2},1,4\rangle 
		-\frac{{4}}{\sqrt{33}}|\frac{1}{2},0,4\rangle  
		+\sqrt{\frac{3}{11}}|\frac{1}{2},1,3\rangle    
	\end{equation}	
	
	\begin{equation} 
		|^{4}G_{11/2};11/2\rangle =
		|\frac{1}{2},1,4\rangle    
	\end{equation}	
	
	Using these $L$–$S$ basis states, we compute the matrix elements of the operators specified in Eqs.~(\ref{eq:LS-operator})–(\ref{eq:spinspin-expect}).  The corresponding expectation values, expressed in the $[{}^{2}G_{J},{}^{4}G_{J}]$ bases, are:\\
	For $J=5/2$,
	\begin{equation}
		\text{$\langle\mathbf{L}.\mathbf{S_l}\rangle$=}-\frac{5}{2},\ \  
		\text{$\langle\mathbf{L}.\mathbf{S_h}\rangle$=}-5,\ \   
		\text{$\langle\mathbf{\hat{B}}\rangle$=}-\frac{5}{7},\ \ 
		\text{$\langle\mathbf{S_l}.\mathbf{S_h}\rangle$=}\frac{1}{2}.
	\end{equation}	
	For $J=7/2$,
	\begin{equation}
		\text{$\langle\mathbf{L}.\mathbf{S_l}\rangle$=}\left[
		\begin{array}{cc}
			\frac{5}{6} & -\frac{\sqrt{35}}{3} \\
			-\frac{\sqrt{35}}{3} & -\frac{4}{3} \\
		\end{array}
		\right],\ \  
		\text{$\langle\mathbf{L}.\mathbf{S_h}\rangle$=}\left[
		\begin{array}{cc}
			-\frac{10}{3} & \frac{\sqrt{35}}{3} \\
			\frac{\sqrt{35}}{3} & -\frac{8}{3} \\
		\end{array}
		\right],\ \   
		\text{$\langle\mathbf{\hat{B}}\rangle$=}\left[
		\begin{array}{cc}
			0 & -\frac{\sqrt{\frac{5}{7}}}{2} \\
			-\frac{\sqrt{\frac{5}{7}}}{2} & \frac{2}{7} \\
		\end{array}
		\right],\ \ 
		\text{$\langle\mathbf{S_l}.\mathbf{S_h}\rangle$=}\left[
		\begin{array}{cc}
			-1 & 0 \\
			0 & \frac{1}{2} \\
		\end{array}
		\right].
	\end{equation}	
	For $J=9/2$,
	\begin{equation}
		\text{$\langle\mathbf{L}.\mathbf{S_l}\rangle$=}\left[
		\begin{array}{cc}
			-\frac{2}{3} & -\frac{2 \sqrt{11}}{3} \\
			-\frac{2 \sqrt{11}}{3} & \frac{1}{6} \\
		\end{array}
		\right],\ \  
		\text{$\langle\mathbf{L}.\mathbf{S_h}\rangle$=}\left[
		\begin{array}{cc}
			\frac{8}{3} & \frac{2 \sqrt{11}}{3} \\
			\frac{2 \sqrt{11}}{3} & \frac{1}{3} \\
		\end{array}
		\right],\ \   
		\text{$\langle\mathbf{\hat{B}}\rangle$=}\left[
		\begin{array}{cc}
			0 & \frac{1}{\sqrt{11}} \\
			\frac{1}{\sqrt{11}} & \frac{7}{11} \\
		\end{array}
		\right],\ \ 
		\text{$\langle\mathbf{S_l}.\mathbf{S_h}\rangle$=}\left[
		\begin{array}{cc}
			-1 & 0 \\
			0 & \frac{1}{2} \\
		\end{array}
		\right].
	\end{equation}	
	For $J=11/2$,
	\begin{equation}
		\text{$\langle\mathbf{L}.\mathbf{S_l}\rangle$=}2,\ \  
		\text{$\langle\mathbf{L}.\mathbf{S_h}\rangle$=}4,\ \   
		\text{$\langle\mathbf{\hat{B}}\rangle$=}-\frac{4}{11},\ \ 
		\text{$\langle\mathbf{S_l}.\mathbf{S_h}\rangle$=}\frac{1}{2}.
	\end{equation}	
	Diagonalization of $\langle\mathbf{L}\cdot\mathbf{S}_l\rangle$ for $J=\frac{7}{2},\frac{9}{2}$ yields the $|J, j \rangle$ basis corresponding to the eigenvalues $\lambda=-\tfrac52$ (for $j=\tfrac72$) and $\lambda=2$ (for $j=\tfrac92$):
	\begin{equation}
		\lambda= -\frac{5}{2} :
		|J=\frac{7}{2}, j=7/2 \rangle =
		\frac{1}{3}\sqrt{\frac{7}{3}}|^{2}G_{7/2}   \rangle 
		+\frac{2}{3}\sqrt{\frac{5}{3}}|^{4}G_{7/2}   \rangle,
	\end{equation}
	\begin{equation}
		\lambda= 2 :
		|J=\frac{7}{2}, j=9/2 \rangle =
		-\frac{2}{3}\sqrt{\frac{5}{3}}  |^{2}G_{7/2}   \rangle 
		+\frac{1}{3}\sqrt{\frac{7}{3}}  |^{4}G_{7/2}   \rangle,
	\end{equation}
	\begin{equation}
		\lambda= -\frac{5}{2} :
		|J=\frac{9}{2}, j=7/2 \rangle =
		\frac{4}{3\sqrt{3}}|^{2}G_{9/2}   \rangle 
		+\frac{1}{3}\sqrt{\frac{11}{3}}|^{4}G_{9/2}   \rangle,
	\end{equation}
	\begin{equation}
		\lambda= 2 :
		|J=\frac{9}{2}, j=9/2 \rangle =
		-\frac{1}{3}\sqrt{\frac{11}{3}}  |^{2}G_{9/2}   \rangle 
		+\frac{4}{3\sqrt{3}}  |^{4}G_{9/2}   \rangle,
	\end{equation}
	whereas the bases for $J=\frac{5}{2}, \frac{11}{2}$ remain unmixed:
	\begin{equation}
		|J=\frac{5}{2}, j=7/2 \rangle =
		|^{4}G_{5/2}   \rangle ,
	\end{equation}
	\begin{equation}
		|J=\frac{11}{2}, j=9/2 \rangle =
		|^{4}G_{11/2}   \rangle .
	\end{equation}
	Expectation values of the mass–splitting operators obtained using these $|J,j\rangle$ basis are tabulated in Table \ref{table0.2}.

	\section*{APPENDIX B}

\end{widetext}

\bibliography{apssamp}

\begin{thebibliography}{70}%
\makeatletter
\providecommand \@ifxundefined [1]{%
 \@ifx{#1\undefined}
}%
\providecommand \@ifnum [1]{%
 \ifnum #1\expandafter \@firstoftwo
 \else \expandafter \@secondoftwo
 \fi
}%
\providecommand \@ifx [1]{%
 \ifx #1\expandafter \@firstoftwo
 \else \expandafter \@secondoftwo
 \fi
}%
\providecommand \natexlab [1]{#1}%
\providecommand \enquote  [1]{``#1''}%
\providecommand \bibnamefont  [1]{#1}%
\providecommand \bibfnamefont [1]{#1}%
\providecommand \citenamefont [1]{#1}%
\providecommand \href@noop [0]{\@secondoftwo}%
\providecommand \href [0]{\begingroup \@sanitize@url \@href}%
\providecommand \@href[1]{\@@startlink{#1}\@@href}%
\providecommand \@@href[1]{\endgroup#1\@@endlink}%
\providecommand \@sanitize@url [0]{\catcode `\\12\catcode `\$12\catcode
  `\&12\catcode `\#12\catcode `\^12\catcode `\_12\catcode `\%12\relax}%
\providecommand \@@startlink[1]{}%
\providecommand \@@endlink[0]{}%
\providecommand \url  [0]{\begingroup\@sanitize@url \@url }%
\providecommand \@url [1]{\endgroup\@href {#1}{\urlprefix }}%
\providecommand \urlprefix  [0]{URL }%
\providecommand \Eprint [0]{\href }%
\providecommand \doibase [0]{https://doi.org/}%
\providecommand \selectlanguage [0]{\@gobble}%
\providecommand \bibinfo  [0]{\@secondoftwo}%
\providecommand \bibfield  [0]{\@secondoftwo}%
\providecommand \translation [1]{[#1]}%
\providecommand \BibitemOpen [0]{}%
\providecommand \bibitemStop [0]{}%
\providecommand \bibitemNoStop [0]{.\EOS\space}%
\providecommand \EOS [0]{\spacefactor3000\relax}%
\providecommand \BibitemShut  [1]{\csname bibitem#1\endcsname}%
\let\auto@bib@innerbib\@empty
\bibitem [{\citenamefont {Navas}\ \emph {et~al.}(2024)\citenamefont {Navas}
  \emph {et~al.}}]{PDG2024}%
  \BibitemOpen
  \bibfield  {author} {\bibinfo {author} {\bibfnamefont {S.}~\bibnamefont
  {Navas}} \emph {et~al.} (\bibinfo {collaboration} {Particle Data Group}),\
  }\bibfield  {title} {\bibinfo {title} {{Review of particle physics}},\ }\href
  {https://doi.org/10.1103/PhysRevD.110.030001} {\bibfield  {journal} {\bibinfo
   {journal} {Phys. Rev. D}\ }\textbf {\bibinfo {volume} {110}},\ \bibinfo
  {pages} {030001} (\bibinfo {year} {2024})}\BibitemShut {NoStop}%
\bibitem [{\citenamefont {Abazov}\ \emph {et~al.}(2007)\citenamefont {Abazov}
  \emph {et~al.}}]{D0:2007vzd}%
  \BibitemOpen
  \bibfield  {author} {\bibinfo {author} {\bibfnamefont {V.~M.}\ \bibnamefont
  {Abazov}} \emph {et~al.} (\bibinfo {collaboration} {D0}),\ }\bibfield
  {title} {\bibinfo {title} {{Observation and Properties of $L = 1 B_{1}$ and
  $B^*_2$ Mesons}},\ }\href {https://doi.org/10.1103/PhysRevLett.99.172001}
  {\bibfield  {journal} {\bibinfo  {journal} {Phys. Rev. Lett.}\ }\textbf
  {\bibinfo {volume} {99}},\ \bibinfo {pages} {172001} (\bibinfo {year}
  {2007})},\ \Eprint {https://arxiv.org/abs/0705.3229} {arXiv:0705.3229
  [hep-ex]} \BibitemShut {NoStop}%
\bibitem [{\citenamefont {Aaltonen}\ \emph {et~al.}(2009)\citenamefont
  {Aaltonen} \emph {et~al.}}]{CDF:2008qzb}%
  \BibitemOpen
  \bibfield  {author} {\bibinfo {author} {\bibfnamefont {T.}~\bibnamefont
  {Aaltonen}} \emph {et~al.} (\bibinfo {collaboration} {CDF}),\ }\bibfield
  {title} {\bibinfo {title} {{Measurement of Resonance Parameters of Orbitally
  Excited Narrow $B^0$ Mesons}},\ }\href
  {https://doi.org/10.1103/PhysRevLett.102.102003} {\bibfield  {journal}
  {\bibinfo  {journal} {Phys. Rev. Lett.}\ }\textbf {\bibinfo {volume} {102}},\
  \bibinfo {pages} {102003} (\bibinfo {year} {2009})},\ \Eprint
  {https://arxiv.org/abs/0809.5007} {arXiv:0809.5007 [hep-ex]} \BibitemShut
  {NoStop}%
\bibitem [{\citenamefont {Aaltonen}\ \emph {et~al.}(2008)\citenamefont
  {Aaltonen} \emph {et~al.}}]{CDF:2007avt}%
  \BibitemOpen
  \bibfield  {author} {\bibinfo {author} {\bibfnamefont {T.}~\bibnamefont
  {Aaltonen}} \emph {et~al.} (\bibinfo {collaboration} {CDF}),\ }\bibfield
  {title} {\bibinfo {title} {{Observation of orbitally excited $B_s$ mesons}},\
  }\href {https://doi.org/10.1103/PhysRevLett.100.082001} {\bibfield  {journal}
  {\bibinfo  {journal} {Phys. Rev. Lett.}\ }\textbf {\bibinfo {volume} {100}},\
  \bibinfo {pages} {082001} (\bibinfo {year} {2008})},\ \Eprint
  {https://arxiv.org/abs/0710.4199} {arXiv:0710.4199 [hep-ex]} \BibitemShut
  {NoStop}%
\bibitem [{\citenamefont {Aaij}\ \emph {et~al.}(2013)\citenamefont {Aaij} \emph
  {et~al.}}]{LHCb:2012iuq}%
  \BibitemOpen
  \bibfield  {author} {\bibinfo {author} {\bibfnamefont {R.}~\bibnamefont
  {Aaij}} \emph {et~al.} (\bibinfo {collaboration} {LHCb}),\ }\bibfield
  {title} {\bibinfo {title} {{First observation of the decay $B_{s2}^*(5840)^0
  \to B^{*+} K^-$ and studies of excited $B^0_s$ mesons}},\ }\href
  {https://doi.org/10.1103/PhysRevLett.110.151803} {\bibfield  {journal}
  {\bibinfo  {journal} {Phys. Rev. Lett.}\ }\textbf {\bibinfo {volume} {110}},\
  \bibinfo {pages} {151803} (\bibinfo {year} {2013})},\ \Eprint
  {https://arxiv.org/abs/1211.5994} {arXiv:1211.5994 [hep-ex]} \BibitemShut
  {NoStop}%
\bibitem [{\citenamefont {Abazov}\ \emph {et~al.}(2008)\citenamefont {Abazov}
  \emph {et~al.}}]{D0:2007die}%
  \BibitemOpen
  \bibfield  {author} {\bibinfo {author} {\bibfnamefont {V.~M.}\ \bibnamefont
  {Abazov}} \emph {et~al.} (\bibinfo {collaboration} {D0}),\ }\bibfield
  {title} {\bibinfo {title} {{Observation and properties of the orbitally
  excited B*(s2) meson}},\ }\href
  {https://doi.org/10.1103/PhysRevLett.100.082002} {\bibfield  {journal}
  {\bibinfo  {journal} {Phys. Rev. Lett.}\ }\textbf {\bibinfo {volume} {100}},\
  \bibinfo {pages} {082002} (\bibinfo {year} {2008})},\ \Eprint
  {https://arxiv.org/abs/0711.0319} {arXiv:0711.0319 [hep-ex]} \BibitemShut
  {NoStop}%
\bibitem [{\citenamefont {Aaltonen}\ \emph {et~al.}(2014)\citenamefont
  {Aaltonen} \emph {et~al.}}]{CDF:2013www}%
  \BibitemOpen
  \bibfield  {author} {\bibinfo {author} {\bibfnamefont {T.~A.}\ \bibnamefont
  {Aaltonen}} \emph {et~al.} (\bibinfo {collaboration} {CDF}),\ }\bibfield
  {title} {\bibinfo {title} {{Study of Orbitally Excited $B$ Mesons and
  Evidence for a New $B\pi$ Resonance}},\ }\href
  {https://doi.org/10.1103/PhysRevD.90.012013} {\bibfield  {journal} {\bibinfo
  {journal} {Phys. Rev. D}\ }\textbf {\bibinfo {volume} {90}},\ \bibinfo
  {pages} {012013} (\bibinfo {year} {2014})},\ \Eprint
  {https://arxiv.org/abs/1309.5961} {arXiv:1309.5961 [hep-ex]} \BibitemShut
  {NoStop}%
\bibitem [{\citenamefont {Aaij}\ \emph {et~al.}(2015)\citenamefont {Aaij} \emph
  {et~al.}}]{LHCb:2015aaf}%
  \BibitemOpen
  \bibfield  {author} {\bibinfo {author} {\bibfnamefont {R.}~\bibnamefont
  {Aaij}} \emph {et~al.} (\bibinfo {collaboration} {LHCb}),\ }\bibfield
  {title} {\bibinfo {title} {{Precise measurements of the properties of the
  $B_1(5721)^{0,+}$ and $B^\ast_2(5747)^{0,+}$ states and observation of
  $B^{+,0}\pi^{-,+}$ mass structures}},\ }\href
  {https://doi.org/10.1007/JHEP04(2015)024} {\bibfield  {journal} {\bibinfo
  {journal} {JHEP}\ }\textbf {\bibinfo {volume} {04}},\ \bibinfo {pages}
  {024}},\ \Eprint {https://arxiv.org/abs/1502.02638} {arXiv:1502.02638
  [hep-ex]} \BibitemShut {NoStop}%
\bibitem [{\citenamefont {Aaij}\ \emph {et~al.}(2021)\citenamefont {Aaij} \emph
  {et~al.}}]{LHCb:2020pet}%
  \BibitemOpen
  \bibfield  {author} {\bibinfo {author} {\bibfnamefont {R.}~\bibnamefont
  {Aaij}} \emph {et~al.} (\bibinfo {collaboration} {LHCb}),\ }\bibfield
  {title} {\bibinfo {title} {{Observation of new excited ${B} ^0_{s} $
  states}},\ }\href {https://doi.org/10.1140/epjc/s10052-021-09305-3}
  {\bibfield  {journal} {\bibinfo  {journal} {Eur. Phys. J. C}\ }\textbf
  {\bibinfo {volume} {81}},\ \bibinfo {pages} {601} (\bibinfo {year} {2021})},\
  \Eprint {https://arxiv.org/abs/2010.15931} {arXiv:2010.15931 [hep-ex]}
  \BibitemShut {NoStop}%
\bibitem [{\citenamefont {Aaij}\ \emph {et~al.}(2017)\citenamefont {Aaij} \emph
  {et~al.}}]{LHCb:2017iph}%
  \BibitemOpen
  \bibfield  {author} {\bibinfo {author} {\bibfnamefont {R.}~\bibnamefont
  {Aaij}} \emph {et~al.} (\bibinfo {collaboration} {LHCb}),\ }\bibfield
  {title} {\bibinfo {title} {{Observation of the doubly charmed baryon
  $\Xi_{cc}^{++}$}},\ }\href {https://doi.org/10.1103/PhysRevLett.119.112001}
  {\bibfield  {journal} {\bibinfo  {journal} {Phys. Rev. Lett.}\ }\textbf
  {\bibinfo {volume} {119}},\ \bibinfo {pages} {112001} (\bibinfo {year}
  {2017})},\ \Eprint {https://arxiv.org/abs/1707.01621} {arXiv:1707.01621
  [hep-ex]} \BibitemShut {NoStop}%
\bibitem [{\citenamefont {Ebert}\ \emph {et~al.}(2010)\citenamefont {Ebert},
  \citenamefont {Faustov},\ and\ \citenamefont {Galkin}}]{Ebert:2009ua}%
  \BibitemOpen
  \bibfield  {author} {\bibinfo {author} {\bibfnamefont {D.}~\bibnamefont
  {Ebert}}, \bibinfo {author} {\bibfnamefont {R.~N.}\ \bibnamefont {Faustov}},\
  and\ \bibinfo {author} {\bibfnamefont {V.~O.}\ \bibnamefont {Galkin}},\
  }\bibfield  {title} {\bibinfo {title} {{Heavy-light meson spectroscopy and
  Regge trajectories in the relativistic quark model}},\ }\href
  {https://doi.org/10.1140/epjc/s10052-010-1233-6} {\bibfield  {journal}
  {\bibinfo  {journal} {Eur. Phys. J. C}\ }\textbf {\bibinfo {volume} {66}},\
  \bibinfo {pages} {197} (\bibinfo {year} {2010})},\ \Eprint
  {https://arxiv.org/abs/0910.5612} {arXiv:0910.5612 [hep-ph]} \BibitemShut
  {NoStop}%
\bibitem [{\citenamefont {Godfrey}\ \emph {et~al.}(2016)\citenamefont
  {Godfrey}, \citenamefont {Moats},\ and\ \citenamefont
  {Swanson}}]{Godfrey:2016nwn}%
  \BibitemOpen
  \bibfield  {author} {\bibinfo {author} {\bibfnamefont {S.}~\bibnamefont
  {Godfrey}}, \bibinfo {author} {\bibfnamefont {K.}~\bibnamefont {Moats}},\
  and\ \bibinfo {author} {\bibfnamefont {E.~S.}\ \bibnamefont {Swanson}},\
  }\bibfield  {title} {\bibinfo {title} {{$B$ and $B_s$ Meson Spectroscopy}},\
  }\href {https://doi.org/10.1103/PhysRevD.94.054025} {\bibfield  {journal}
  {\bibinfo  {journal} {Phys. Rev. D}\ }\textbf {\bibinfo {volume} {94}},\
  \bibinfo {pages} {054025} (\bibinfo {year} {2016})},\ \Eprint
  {https://arxiv.org/abs/1607.02169} {arXiv:1607.02169 [hep-ph]} \BibitemShut
  {NoStop}%
\bibitem [{\citenamefont {Sun}\ \emph {et~al.}(2014)\citenamefont {Sun},
  \citenamefont {Song}, \citenamefont {Chen}, \citenamefont {Liu},\ and\
  \citenamefont {Zhu}}]{Sun:2014wea}%
  \BibitemOpen
  \bibfield  {author} {\bibinfo {author} {\bibfnamefont {Y.}~\bibnamefont
  {Sun}}, \bibinfo {author} {\bibfnamefont {Q.-T.}\ \bibnamefont {Song}},
  \bibinfo {author} {\bibfnamefont {D.-Y.}\ \bibnamefont {Chen}}, \bibinfo
  {author} {\bibfnamefont {X.}~\bibnamefont {Liu}},\ and\ \bibinfo {author}
  {\bibfnamefont {S.-L.}\ \bibnamefont {Zhu}},\ }\bibfield  {title} {\bibinfo
  {title} {{Higher bottom and bottom-strange mesons}},\ }\href
  {https://doi.org/10.1103/PhysRevD.89.054026} {\bibfield  {journal} {\bibinfo
  {journal} {Phys. Rev. D}\ }\textbf {\bibinfo {volume} {89}},\ \bibinfo
  {pages} {054026} (\bibinfo {year} {2014})},\ \Eprint
  {https://arxiv.org/abs/1401.1595} {arXiv:1401.1595 [hep-ph]} \BibitemShut
  {NoStop}%
\bibitem [{\citenamefont {li}\ \emph {et~al.}(2021)\citenamefont {li},
  \citenamefont {Ni},\ and\ \citenamefont {Zhong}}]{li:2021hss}%
  \BibitemOpen
  \bibfield  {author} {\bibinfo {author} {\bibfnamefont {Q.}~\bibnamefont
  {li}}, \bibinfo {author} {\bibfnamefont {R.-H.}\ \bibnamefont {Ni}},\ and\
  \bibinfo {author} {\bibfnamefont {X.-H.}\ \bibnamefont {Zhong}},\ }\bibfield
  {title} {\bibinfo {title} {{Towards establishing an abundant $B$ and $B_s$
  spectrum up to the second orbital excitations}},\ }\href
  {https://doi.org/10.1103/PhysRevD.103.116010} {\bibfield  {journal} {\bibinfo
   {journal} {Phys. Rev. D}\ }\textbf {\bibinfo {volume} {103}},\ \bibinfo
  {pages} {116010} (\bibinfo {year} {2021})},\ \Eprint
  {https://arxiv.org/abs/2102.03694} {arXiv:2102.03694 [hep-ph]} \BibitemShut
  {NoStop}%
\bibitem [{\citenamefont {Zeng}\ \emph {et~al.}(1995)\citenamefont {Zeng},
  \citenamefont {Van~Orden},\ and\ \citenamefont {Roberts}}]{Zeng:1994vj}%
  \BibitemOpen
  \bibfield  {author} {\bibinfo {author} {\bibfnamefont {J.}~\bibnamefont
  {Zeng}}, \bibinfo {author} {\bibfnamefont {J.~W.}\ \bibnamefont
  {Van~Orden}},\ and\ \bibinfo {author} {\bibfnamefont {W.}~\bibnamefont
  {Roberts}},\ }\bibfield  {title} {\bibinfo {title} {{Heavy mesons in a
  relativistic model}},\ }\href {https://doi.org/10.1103/PhysRevD.52.5229}
  {\bibfield  {journal} {\bibinfo  {journal} {Phys. Rev. D}\ }\textbf {\bibinfo
  {volume} {52}},\ \bibinfo {pages} {5229} (\bibinfo {year} {1995})},\ \Eprint
  {https://arxiv.org/abs/hep-ph/9412269} {arXiv:hep-ph/9412269} \BibitemShut
  {NoStop}%
\bibitem [{\citenamefont {Shah}\ \emph
  {et~al.}(2016{\natexlab{a}})\citenamefont {Shah}, \citenamefont {Patel},\
  and\ \citenamefont {Vinodkumar}}]{Shah:2016mgq}%
  \BibitemOpen
  \bibfield  {author} {\bibinfo {author} {\bibfnamefont {M.}~\bibnamefont
  {Shah}}, \bibinfo {author} {\bibfnamefont {B.}~\bibnamefont {Patel}},\ and\
  \bibinfo {author} {\bibfnamefont {P.~C.}\ \bibnamefont {Vinodkumar}},\
  }\bibfield  {title} {\bibinfo {title} {{Spectroscopy and flavor changing
  decays of B , B$_s$ mesons in a Dirac formalism}},\ }\href
  {https://doi.org/10.1103/PhysRevD.93.094028} {\bibfield  {journal} {\bibinfo
  {journal} {Phys. Rev. D}\ }\textbf {\bibinfo {volume} {93}},\ \bibinfo
  {pages} {094028} (\bibinfo {year} {2016}{\natexlab{a}})}\BibitemShut
  {NoStop}%
\bibitem [{\citenamefont {Devlani}\ and\ \citenamefont
  {Rai}(2011)}]{Devlani:2011zz}%
  \BibitemOpen
  \bibfield  {author} {\bibinfo {author} {\bibfnamefont {N.}~\bibnamefont
  {Devlani}}\ and\ \bibinfo {author} {\bibfnamefont {A.~K.}\ \bibnamefont
  {Rai}},\ }\bibfield  {title} {\bibinfo {title} {{Spectroscopy and decay
  properties of the Ds meson}},\ }\href
  {https://doi.org/10.1103/PhysRevD.84.074030} {\bibfield  {journal} {\bibinfo
  {journal} {Phys. Rev. D}\ }\textbf {\bibinfo {volume} {84}},\ \bibinfo
  {pages} {074030} (\bibinfo {year} {2011})}\BibitemShut {NoStop}%
\bibitem [{\citenamefont {L\"u}\ \emph {et~al.}(2017)\citenamefont {L\"u},
  \citenamefont {Wang}, \citenamefont {Xiao},\ and\ \citenamefont
  {Zhong}}]{Lu:2017meb}%
  \BibitemOpen
  \bibfield  {author} {\bibinfo {author} {\bibfnamefont {Q.-F.}\ \bibnamefont
  {L\"u}}, \bibinfo {author} {\bibfnamefont {K.-L.}\ \bibnamefont {Wang}},
  \bibinfo {author} {\bibfnamefont {L.-Y.}\ \bibnamefont {Xiao}},\ and\
  \bibinfo {author} {\bibfnamefont {X.-H.}\ \bibnamefont {Zhong}},\ }\bibfield
  {title} {\bibinfo {title} {{Mass spectra and radiative transitions of doubly
  heavy baryons in a relativized quark model}},\ }\href
  {https://doi.org/10.1103/PhysRevD.96.114006} {\bibfield  {journal} {\bibinfo
  {journal} {Phys. Rev. D}\ }\textbf {\bibinfo {volume} {96}},\ \bibinfo
  {pages} {114006} (\bibinfo {year} {2017})},\ \Eprint
  {https://arxiv.org/abs/1708.04468} {arXiv:1708.04468 [hep-ph]} \BibitemShut
  {NoStop}%
\bibitem [{\citenamefont {Ebert}\ \emph {et~al.}(2002)\citenamefont {Ebert},
  \citenamefont {Faustov}, \citenamefont {Galkin},\ and\ \citenamefont
  {Martynenko}}]{Ebert:2002ig}%
  \BibitemOpen
  \bibfield  {author} {\bibinfo {author} {\bibfnamefont {D.}~\bibnamefont
  {Ebert}}, \bibinfo {author} {\bibfnamefont {R.~N.}\ \bibnamefont {Faustov}},
  \bibinfo {author} {\bibfnamefont {V.~O.}\ \bibnamefont {Galkin}},\ and\
  \bibinfo {author} {\bibfnamefont {A.~P.}\ \bibnamefont {Martynenko}},\
  }\bibfield  {title} {\bibinfo {title} {{Mass spectra of doubly heavy baryons
  in the relativistic quark model}},\ }\href
  {https://doi.org/10.1103/PhysRevD.66.014008} {\bibfield  {journal} {\bibinfo
  {journal} {Phys. Rev. D}\ }\textbf {\bibinfo {volume} {66}},\ \bibinfo
  {pages} {014008} (\bibinfo {year} {2002})},\ \Eprint
  {https://arxiv.org/abs/hep-ph/0201217} {arXiv:hep-ph/0201217} \BibitemShut
  {NoStop}%
\bibitem [{\citenamefont {Gershtein}\ \emph {et~al.}(2000)\citenamefont
  {Gershtein}, \citenamefont {Kiselev}, \citenamefont {Likhoded},\ and\
  \citenamefont {Onishchenko}}]{Gershtein:2000nx}%
  \BibitemOpen
  \bibfield  {author} {\bibinfo {author} {\bibfnamefont {S.~S.}\ \bibnamefont
  {Gershtein}}, \bibinfo {author} {\bibfnamefont {V.~V.}\ \bibnamefont
  {Kiselev}}, \bibinfo {author} {\bibfnamefont {A.~K.}\ \bibnamefont
  {Likhoded}},\ and\ \bibinfo {author} {\bibfnamefont {A.~I.}\ \bibnamefont
  {Onishchenko}},\ }\bibfield  {title} {\bibinfo {title} {{Spectroscopy of
  doubly heavy baryons}},\ }\href {https://doi.org/10.1103/PhysRevD.62.054021}
  {\bibfield  {journal} {\bibinfo  {journal} {Phys. Rev. D}\ }\textbf {\bibinfo
  {volume} {62}},\ \bibinfo {pages} {054021} (\bibinfo {year}
  {2000})}\BibitemShut {NoStop}%
\bibitem [{\citenamefont {Yoshida}\ \emph {et~al.}(2015)\citenamefont
  {Yoshida}, \citenamefont {Hiyama}, \citenamefont {Hosaka}, \citenamefont
  {Oka},\ and\ \citenamefont {Sadato}}]{Yoshida:2015tia}%
  \BibitemOpen
  \bibfield  {author} {\bibinfo {author} {\bibfnamefont {T.}~\bibnamefont
  {Yoshida}}, \bibinfo {author} {\bibfnamefont {E.}~\bibnamefont {Hiyama}},
  \bibinfo {author} {\bibfnamefont {A.}~\bibnamefont {Hosaka}}, \bibinfo
  {author} {\bibfnamefont {M.}~\bibnamefont {Oka}},\ and\ \bibinfo {author}
  {\bibfnamefont {K.}~\bibnamefont {Sadato}},\ }\bibfield  {title} {\bibinfo
  {title} {{Spectrum of heavy baryons in the quark model}},\ }\href
  {https://doi.org/10.1103/PhysRevD.92.114029} {\bibfield  {journal} {\bibinfo
  {journal} {Phys. Rev. D}\ }\textbf {\bibinfo {volume} {92}},\ \bibinfo
  {pages} {114029} (\bibinfo {year} {2015})},\ \Eprint
  {https://arxiv.org/abs/1510.01067} {arXiv:1510.01067 [hep-ph]} \BibitemShut
  {NoStop}%
\bibitem [{\citenamefont {Giannuzzi}(2009)}]{Giannuzzi:2009gh}%
  \BibitemOpen
  \bibfield  {author} {\bibinfo {author} {\bibfnamefont {F.}~\bibnamefont
  {Giannuzzi}},\ }\bibfield  {title} {\bibinfo {title} {{Doubly heavy baryons
  in a Salpeter model with AdS/QCD inspired potential}},\ }\href
  {https://doi.org/10.1103/PhysRevD.79.094002} {\bibfield  {journal} {\bibinfo
  {journal} {Phys. Rev. D}\ }\textbf {\bibinfo {volume} {79}},\ \bibinfo
  {pages} {094002} (\bibinfo {year} {2009})},\ \Eprint
  {https://arxiv.org/abs/0902.4624} {arXiv:0902.4624 [hep-ph]} \BibitemShut
  {NoStop}%
\bibitem [{\citenamefont {Rai}\ \emph {et~al.}(2008)\citenamefont {Rai},
  \citenamefont {Patel},\ and\ \citenamefont {Vinodkumar}}]{Rai:2008sc}%
  \BibitemOpen
  \bibfield  {author} {\bibinfo {author} {\bibfnamefont {A.~K.}\ \bibnamefont
  {Rai}}, \bibinfo {author} {\bibfnamefont {B.}~\bibnamefont {Patel}},\ and\
  \bibinfo {author} {\bibfnamefont {P.~C.}\ \bibnamefont {Vinodkumar}},\
  }\bibfield  {title} {\bibinfo {title} {{Properties of $Q \bar{Q}$ mesons in
  non-relativistic QCD formalism}},\ }\href
  {https://doi.org/10.1103/PhysRevC.78.055202} {\bibfield  {journal} {\bibinfo
  {journal} {Phys. Rev. C}\ }\textbf {\bibinfo {volume} {78}},\ \bibinfo
  {pages} {055202} (\bibinfo {year} {2008})},\ \Eprint
  {https://arxiv.org/abs/0810.1832} {arXiv:0810.1832 [hep-ph]} \BibitemShut
  {NoStop}%
\bibitem [{\citenamefont {Kher}\ \emph {et~al.}(2017)\citenamefont {Kher},
  \citenamefont {Devlani},\ and\ \citenamefont {Rai}}]{Kher:2017mky}%
  \BibitemOpen
  \bibfield  {author} {\bibinfo {author} {\bibfnamefont {V.}~\bibnamefont
  {Kher}}, \bibinfo {author} {\bibfnamefont {N.}~\bibnamefont {Devlani}},\ and\
  \bibinfo {author} {\bibfnamefont {A.~K.}\ \bibnamefont {Rai}},\ }\bibfield
  {title} {\bibinfo {title} {{Spectroscopy, Decay properties and Regge
  trajectories of the $B$ and $B_s$ mesons}},\ }\href
  {https://doi.org/10.1088/1674-1137/41/9/093101} {\bibfield  {journal}
  {\bibinfo  {journal} {Chin. Phys. C}\ }\textbf {\bibinfo {volume} {41}},\
  \bibinfo {pages} {093101} (\bibinfo {year} {2017})},\ \Eprint
  {https://arxiv.org/abs/1705.08248} {arXiv:1705.08248 [hep-ph]} \BibitemShut
  {NoStop}%
\bibitem [{\citenamefont {Yazarloo}\ and\ \citenamefont
  {Mehraban}(2016)}]{Yazarloo:2016luc}%
  \BibitemOpen
  \bibfield  {author} {\bibinfo {author} {\bibfnamefont {B.~H.}\ \bibnamefont
  {Yazarloo}}\ and\ \bibinfo {author} {\bibfnamefont {H.}~\bibnamefont
  {Mehraban}},\ }\bibfield  {title} {\bibinfo {title} {{Study of B and B$_s$
  mesons with a Coulomb plus exponential type potential}},\ }\href
  {https://doi.org/10.1209/0295-5075/116/31004} {\bibfield  {journal} {\bibinfo
   {journal} {EPL}\ }\textbf {\bibinfo {volume} {116}},\ \bibinfo {pages}
  {31004} (\bibinfo {year} {2016})}\BibitemShut {NoStop}%
\bibitem [{\citenamefont {Patel}\ \emph {et~al.}(2024)\citenamefont {Patel},
  \citenamefont {Chaturvedi},\ and\ \citenamefont {Rai}}]{Patel:2024cng}%
  \BibitemOpen
  \bibfield  {author} {\bibinfo {author} {\bibfnamefont {V.}~\bibnamefont
  {Patel}}, \bibinfo {author} {\bibfnamefont {R.}~\bibnamefont {Chaturvedi}},\
  and\ \bibinfo {author} {\bibfnamefont {A.~K.}\ \bibnamefont {Rai}},\
  }\bibfield  {title} {\bibinfo {title} {{Spectroscopic properties of $ B $ and
  $ B_s $ meson using screened potential}},\ }\href
  {https://doi.org/10.1007/s12648-023-03048-5} {\bibfield  {journal} {\bibinfo
  {journal} {Indian J. Phys.}\ }\textbf {\bibinfo {volume} {98}},\ \bibinfo
  {pages} {2961} (\bibinfo {year} {2024})}\BibitemShut {NoStop}%
\bibitem [{\citenamefont {Eakins}\ and\ \citenamefont
  {Roberts}(2012)}]{Eakins:2012jk}%
  \BibitemOpen
  \bibfield  {author} {\bibinfo {author} {\bibfnamefont {B.}~\bibnamefont
  {Eakins}}\ and\ \bibinfo {author} {\bibfnamefont {W.}~\bibnamefont
  {Roberts}},\ }\bibfield  {title} {\bibinfo {title} {{Symmetries and
  Systematics of Doubly Heavy Hadrons}},\ }\href
  {https://doi.org/10.1142/S0217751X1250039X} {\bibfield  {journal} {\bibinfo
  {journal} {Int. J. Mod. Phys. A}\ }\textbf {\bibinfo {volume} {27}},\
  \bibinfo {pages} {1250039} (\bibinfo {year} {2012})},\ \Eprint
  {https://arxiv.org/abs/1201.4885} {arXiv:1201.4885 [nucl-th]} \BibitemShut
  {NoStop}%
\bibitem [{\citenamefont {Chen}\ \emph {et~al.}(2022)\citenamefont {Chen},
  \citenamefont {Luo}, \citenamefont {Wei},\ and\ \citenamefont
  {Liu}}]{PhysRevD.105.074014}%
  \BibitemOpen
  \bibfield  {author} {\bibinfo {author} {\bibfnamefont {B.}~\bibnamefont
  {Chen}}, \bibinfo {author} {\bibfnamefont {S.-Q.}\ \bibnamefont {Luo}},
  \bibinfo {author} {\bibfnamefont {K.-W.}\ \bibnamefont {Wei}},\ and\ \bibinfo
  {author} {\bibfnamefont {X.}~\bibnamefont {Liu}},\ }\bibfield  {title}
  {\bibinfo {title} {$b$-hadron spectroscopy study based on the similarity of
  double bottom baryon and bottom meson},\ }\href
  {https://doi.org/10.1103/PhysRevD.105.074014} {\bibfield  {journal} {\bibinfo
   {journal} {Phys. Rev. D}\ }\textbf {\bibinfo {volume} {105}},\ \bibinfo
  {pages} {074014} (\bibinfo {year} {2022})}\BibitemShut {NoStop}%
\bibitem [{\citenamefont {Kiselev}\ \emph {et~al.}(2002)\citenamefont
  {Kiselev}, \citenamefont {Likhoded}, \citenamefont {Pakhomova},\ and\
  \citenamefont {Saleev}}]{Kiselev:2002iy}%
  \BibitemOpen
  \bibfield  {author} {\bibinfo {author} {\bibfnamefont {V.~V.}\ \bibnamefont
  {Kiselev}}, \bibinfo {author} {\bibfnamefont {A.~K.}\ \bibnamefont
  {Likhoded}}, \bibinfo {author} {\bibfnamefont {O.~N.}\ \bibnamefont
  {Pakhomova}},\ and\ \bibinfo {author} {\bibfnamefont {V.~A.}\ \bibnamefont
  {Saleev}},\ }\bibfield  {title} {\bibinfo {title} {{Mass spectra of doubly
  heavy Omega $Q Q^\prime$ baryons}},\ }\href
  {https://doi.org/10.1103/PhysRevD.66.034030} {\bibfield  {journal} {\bibinfo
  {journal} {Phys. Rev. D}\ }\textbf {\bibinfo {volume} {66}},\ \bibinfo
  {pages} {034030} (\bibinfo {year} {2002})},\ \Eprint
  {https://arxiv.org/abs/hep-ph/0206140} {arXiv:hep-ph/0206140} \BibitemShut
  {NoStop}%
\bibitem [{\citenamefont {Roberts}\ and\ \citenamefont
  {Pervin}(2008)}]{Roberts:2007ni}%
  \BibitemOpen
  \bibfield  {author} {\bibinfo {author} {\bibfnamefont {W.}~\bibnamefont
  {Roberts}}\ and\ \bibinfo {author} {\bibfnamefont {M.}~\bibnamefont
  {Pervin}},\ }\bibfield  {title} {\bibinfo {title} {{Heavy baryons in a quark
  model}},\ }\href {https://doi.org/10.1142/S0217751X08041219} {\bibfield
  {journal} {\bibinfo  {journal} {Int. J. Mod. Phys. A}\ }\textbf {\bibinfo
  {volume} {23}},\ \bibinfo {pages} {2817} (\bibinfo {year} {2008})},\ \Eprint
  {https://arxiv.org/abs/0711.2492} {arXiv:0711.2492 [nucl-th]} \BibitemShut
  {NoStop}%
\bibitem [{\citenamefont {Shah}\ \emph
  {et~al.}(2016{\natexlab{b}})\citenamefont {Shah}, \citenamefont {Thakkar},\
  and\ \citenamefont {Rai}}]{Shah:2016vmd}%
  \BibitemOpen
  \bibfield  {author} {\bibinfo {author} {\bibfnamefont {Z.}~\bibnamefont
  {Shah}}, \bibinfo {author} {\bibfnamefont {K.}~\bibnamefont {Thakkar}},\ and\
  \bibinfo {author} {\bibfnamefont {A.~K.}\ \bibnamefont {Rai}},\ }\bibfield
  {title} {\bibinfo {title} {{Excited State Mass spectra of doubly heavy
  baryons $\Omega_{cc}$, $\Omega_{bb}$ and $\Omega_{bc}$}},\ }\href
  {https://doi.org/10.1140/epjc/s10052-016-4379-z} {\bibfield  {journal}
  {\bibinfo  {journal} {Eur. Phys. J. C}\ }\textbf {\bibinfo {volume} {76}},\
  \bibinfo {pages} {530} (\bibinfo {year} {2016}{\natexlab{b}})},\ \Eprint
  {https://arxiv.org/abs/1609.03030} {arXiv:1609.03030 [hep-ph]} \BibitemShut
  {NoStop}%
\bibitem [{\citenamefont {Shah}\ and\ \citenamefont
  {Rai}(2017)}]{Shah:2017liu}%
  \BibitemOpen
  \bibfield  {author} {\bibinfo {author} {\bibfnamefont {Z.}~\bibnamefont
  {Shah}}\ and\ \bibinfo {author} {\bibfnamefont {A.~K.}\ \bibnamefont {Rai}},\
  }\bibfield  {title} {\bibinfo {title} {{Excited state mass spectra of doubly
  heavy $\Xi$ baryons}},\ }\href
  {https://doi.org/10.1140/epjc/s10052-017-4688-x} {\bibfield  {journal}
  {\bibinfo  {journal} {Eur. Phys. J. C}\ }\textbf {\bibinfo {volume} {77}},\
  \bibinfo {pages} {129} (\bibinfo {year} {2017})},\ \Eprint
  {https://arxiv.org/abs/1702.02726} {arXiv:1702.02726 [hep-ph]} \BibitemShut
  {NoStop}%
\bibitem [{\citenamefont {Karliner}\ and\ \citenamefont
  {Rosner}(2014)}]{Karliner:2014gca}%
  \BibitemOpen
  \bibfield  {author} {\bibinfo {author} {\bibfnamefont {M.}~\bibnamefont
  {Karliner}}\ and\ \bibinfo {author} {\bibfnamefont {J.~L.}\ \bibnamefont
  {Rosner}},\ }\bibfield  {title} {\bibinfo {title} {{Baryons with two heavy
  quarks: Masses, production, decays, and detection}},\ }\href
  {https://doi.org/10.1103/PhysRevD.90.094007} {\bibfield  {journal} {\bibinfo
  {journal} {Phys. Rev. D}\ }\textbf {\bibinfo {volume} {90}},\ \bibinfo
  {pages} {094007} (\bibinfo {year} {2014})},\ \Eprint
  {https://arxiv.org/abs/1408.5877} {arXiv:1408.5877 [hep-ph]} \BibitemShut
  {NoStop}%
\bibitem [{\citenamefont {Lu}\ \emph {et~al.}(2017)\citenamefont {Lu},
  \citenamefont {Anwar},\ and\ \citenamefont {Zou}}]{Lu:2017hma}%
  \BibitemOpen
  \bibfield  {author} {\bibinfo {author} {\bibfnamefont {Y.}~\bibnamefont
  {Lu}}, \bibinfo {author} {\bibfnamefont {M.~N.}\ \bibnamefont {Anwar}},\ and\
  \bibinfo {author} {\bibfnamefont {B.-S.}\ \bibnamefont {Zou}},\ }\bibfield
  {title} {\bibinfo {title} {{How Large is the Contribution of Excited Mesons
  in Coupled-Channel Effects?}},\ }\href
  {https://doi.org/10.1103/PhysRevD.95.034018} {\bibfield  {journal} {\bibinfo
  {journal} {Phys. Rev. D}\ }\textbf {\bibinfo {volume} {95}},\ \bibinfo
  {pages} {034018} (\bibinfo {year} {2017})},\ \Eprint
  {https://arxiv.org/abs/1701.00692} {arXiv:1701.00692 [hep-ph]} \BibitemShut
  {NoStop}%
\bibitem [{\citenamefont {Oudichhya}\ and\ \citenamefont
  {Rai}(2024)}]{Oudichhya:2024hmn}%
  \BibitemOpen
  \bibfield  {author} {\bibinfo {author} {\bibfnamefont {J.}~\bibnamefont
  {Oudichhya}}\ and\ \bibinfo {author} {\bibfnamefont {A.~K.}\ \bibnamefont
  {Rai}},\ }\bibfield  {title} {\bibinfo {title} {{Study of singly bottom and
  doubly heavy baryons within Regge phenomenology}},\ }\href
  {https://doi.org/10.1140/epja/s10050-024-01344-0} {\bibfield  {journal}
  {\bibinfo  {journal} {Eur. Phys. J. A}\ }\textbf {\bibinfo {volume} {60}},\
  \bibinfo {pages} {125} (\bibinfo {year} {2024})}\BibitemShut {NoStop}%
\bibitem [{\citenamefont {Korner}\ \emph {et~al.}(1994)\citenamefont {Korner},
  \citenamefont {Kramer},\ and\ \citenamefont {Pirjol}}]{Korner:1994nh}%
  \BibitemOpen
  \bibfield  {author} {\bibinfo {author} {\bibfnamefont {J.~G.}\ \bibnamefont
  {Korner}}, \bibinfo {author} {\bibfnamefont {M.}~\bibnamefont {Kramer}},\
  and\ \bibinfo {author} {\bibfnamefont {D.}~\bibnamefont {Pirjol}},\
  }\bibfield  {title} {\bibinfo {title} {{Heavy baryons}},\ }\href
  {https://doi.org/10.1016/0146-6410(94)90053-1} {\bibfield  {journal}
  {\bibinfo  {journal} {Prog. Part. Nucl. Phys.}\ }\textbf {\bibinfo {volume}
  {33}},\ \bibinfo {pages} {787} (\bibinfo {year} {1994})},\ \Eprint
  {https://arxiv.org/abs/hep-ph/9406359} {arXiv:hep-ph/9406359} \BibitemShut
  {NoStop}%
\bibitem [{\citenamefont {Cheng}\ and\ \citenamefont
  {Yu}(2017)}]{Cheng:2017oqh}%
  \BibitemOpen
  \bibfield  {author} {\bibinfo {author} {\bibfnamefont {H.-Y.}\ \bibnamefont
  {Cheng}}\ and\ \bibinfo {author} {\bibfnamefont {F.-S.}\ \bibnamefont {Yu}},\
  }\bibfield  {title} {\bibinfo {title} {{Masses of Scalar and Axial-Vector B
  Mesons Revisited}},\ }\href {https://doi.org/10.1140/epjc/s10052-017-5252-4}
  {\bibfield  {journal} {\bibinfo  {journal} {Eur. Phys. J. C}\ }\textbf
  {\bibinfo {volume} {77}},\ \bibinfo {pages} {668} (\bibinfo {year} {2017})},\
  \Eprint {https://arxiv.org/abs/1704.01208} {arXiv:1704.01208 [hep-ph]}
  \BibitemShut {NoStop}%
\bibitem [{\citenamefont {Alhakami}(2021)}]{Alhakami:2020vil}%
  \BibitemOpen
  \bibfield  {author} {\bibinfo {author} {\bibfnamefont {M.~H.}\ \bibnamefont
  {Alhakami}},\ }\bibfield  {title} {\bibinfo {title} {{Predictions for the
  beauty meson spectrum}},\ }\href
  {https://doi.org/10.1103/PhysRevD.103.034009} {\bibfield  {journal} {\bibinfo
   {journal} {Phys. Rev. D}\ }\textbf {\bibinfo {volume} {103}},\ \bibinfo
  {pages} {034009} (\bibinfo {year} {2021})},\ \Eprint
  {https://arxiv.org/abs/2006.16878} {arXiv:2006.16878 [hep-ph]} \BibitemShut
  {NoStop}%
\bibitem [{\citenamefont {Bagan}\ \emph {et~al.}(1993)\citenamefont {Bagan},
  \citenamefont {Chabab},\ and\ \citenamefont {Narison}}]{Bagan:1992za}%
  \BibitemOpen
  \bibfield  {author} {\bibinfo {author} {\bibfnamefont {E.}~\bibnamefont
  {Bagan}}, \bibinfo {author} {\bibfnamefont {M.}~\bibnamefont {Chabab}},\ and\
  \bibinfo {author} {\bibfnamefont {S.}~\bibnamefont {Narison}},\ }\bibfield
  {title} {\bibinfo {title} {{Baryons with two heavy quarks from QCD spectral
  sum rules}},\ }\href {https://doi.org/10.1016/0370-2693(93)90090-5}
  {\bibfield  {journal} {\bibinfo  {journal} {Phys. Lett. B}\ }\textbf
  {\bibinfo {volume} {306}},\ \bibinfo {pages} {350} (\bibinfo {year}
  {1993})}\BibitemShut {NoStop}%
\bibitem [{\citenamefont {Wang}(2010)}]{Wang:2010hs}%
  \BibitemOpen
  \bibfield  {author} {\bibinfo {author} {\bibfnamefont {Z.-G.}\ \bibnamefont
  {Wang}},\ }\bibfield  {title} {\bibinfo {title} {{Analysis of the ${1/2}^+$
  doubly heavy baryon states with QCD sum rules}},\ }\href
  {https://doi.org/10.1140/epja/i2010-11004-3} {\bibfield  {journal} {\bibinfo
  {journal} {Eur. Phys. J. A}\ }\textbf {\bibinfo {volume} {45}},\ \bibinfo
  {pages} {267} (\bibinfo {year} {2010})},\ \Eprint
  {https://arxiv.org/abs/1001.4693} {arXiv:1001.4693 [hep-ph]} \BibitemShut
  {NoStop}%
\bibitem [{\citenamefont {Lang}\ \emph {et~al.}(2015)\citenamefont {Lang},
  \citenamefont {Mohler}, \citenamefont {Prelovsek},\ and\ \citenamefont
  {Woloshyn}}]{Lang:2015hza}%
  \BibitemOpen
  \bibfield  {author} {\bibinfo {author} {\bibfnamefont {C.~B.}\ \bibnamefont
  {Lang}}, \bibinfo {author} {\bibfnamefont {D.}~\bibnamefont {Mohler}},
  \bibinfo {author} {\bibfnamefont {S.}~\bibnamefont {Prelovsek}},\ and\
  \bibinfo {author} {\bibfnamefont {R.~M.}\ \bibnamefont {Woloshyn}},\
  }\bibfield  {title} {\bibinfo {title} {{Predicting positive parity B$_s$
  mesons from lattice QCD}},\ }\href
  {https://doi.org/10.1016/j.physletb.2015.08.038} {\bibfield  {journal}
  {\bibinfo  {journal} {Phys. Lett. B}\ }\textbf {\bibinfo {volume} {750}},\
  \bibinfo {pages} {17} (\bibinfo {year} {2015})},\ \Eprint
  {https://arxiv.org/abs/1501.01646} {arXiv:1501.01646 [hep-lat]} \BibitemShut
  {NoStop}%
\bibitem [{\citenamefont {Gregory}\ \emph {et~al.}(2011)\citenamefont {Gregory}
  \emph {et~al.}}]{Gregory:2010gm}%
  \BibitemOpen
  \bibfield  {author} {\bibinfo {author} {\bibfnamefont {E.~B.}\ \bibnamefont
  {Gregory}} \emph {et~al.},\ }\bibfield  {title} {\bibinfo {title} {{Precise
  $B, B_s$ and $B_c$ meson spectroscopy from full lattice QCD}},\ }\href
  {https://doi.org/10.1103/PhysRevD.83.014506} {\bibfield  {journal} {\bibinfo
  {journal} {Phys. Rev. D}\ }\textbf {\bibinfo {volume} {83}},\ \bibinfo
  {pages} {014506} (\bibinfo {year} {2011})},\ \Eprint
  {https://arxiv.org/abs/1010.3848} {arXiv:1010.3848 [hep-lat]} \BibitemShut
  {NoStop}%
\bibitem [{\citenamefont {Padmanath}\ \emph {et~al.}(2015)\citenamefont
  {Padmanath}, \citenamefont {Edwards}, \citenamefont {Mathur},\ and\
  \citenamefont {Peardon}}]{Padmanath:2015jea}%
  \BibitemOpen
  \bibfield  {author} {\bibinfo {author} {\bibfnamefont {M.}~\bibnamefont
  {Padmanath}}, \bibinfo {author} {\bibfnamefont {R.~G.}\ \bibnamefont
  {Edwards}}, \bibinfo {author} {\bibfnamefont {N.}~\bibnamefont {Mathur}},\
  and\ \bibinfo {author} {\bibfnamefont {M.}~\bibnamefont {Peardon}},\
  }\bibfield  {title} {\bibinfo {title} {{Spectroscopy of doubly-charmed
  baryons from lattice QCD}},\ }\href
  {https://doi.org/10.1103/PhysRevD.91.094502} {\bibfield  {journal} {\bibinfo
  {journal} {Phys. Rev. D}\ }\textbf {\bibinfo {volume} {91}},\ \bibinfo
  {pages} {094502} (\bibinfo {year} {2015})},\ \Eprint
  {https://arxiv.org/abs/1502.01845} {arXiv:1502.01845 [hep-lat]} \BibitemShut
  {NoStop}%
\bibitem [{\citenamefont {Alexandrou}\ \emph {et~al.}(2012)\citenamefont
  {Alexandrou}, \citenamefont {Carbonell}, \citenamefont {Christaras},
  \citenamefont {Drach}, \citenamefont {Gravina},\ and\ \citenamefont
  {Papinutto}}]{Alexandrou:2012xk}%
  \BibitemOpen
  \bibfield  {author} {\bibinfo {author} {\bibfnamefont {C.}~\bibnamefont
  {Alexandrou}}, \bibinfo {author} {\bibfnamefont {J.}~\bibnamefont
  {Carbonell}}, \bibinfo {author} {\bibfnamefont {D.}~\bibnamefont
  {Christaras}}, \bibinfo {author} {\bibfnamefont {V.}~\bibnamefont {Drach}},
  \bibinfo {author} {\bibfnamefont {M.}~\bibnamefont {Gravina}},\ and\ \bibinfo
  {author} {\bibfnamefont {M.}~\bibnamefont {Papinutto}},\ }\bibfield  {title}
  {\bibinfo {title} {{Strange and charm baryon masses with two flavors of
  dynamical twisted mass fermions}},\ }\href
  {https://doi.org/10.1103/PhysRevD.86.114501} {\bibfield  {journal} {\bibinfo
  {journal} {Phys. Rev. D}\ }\textbf {\bibinfo {volume} {86}},\ \bibinfo
  {pages} {114501} (\bibinfo {year} {2012})},\ \Eprint
  {https://arxiv.org/abs/1205.6856} {arXiv:1205.6856 [hep-lat]} \BibitemShut
  {NoStop}%
\bibitem [{\citenamefont {Jakhad}\ \emph {et~al.}(2023)\citenamefont {Jakhad},
  \citenamefont {Oudichhya}, \citenamefont {Gandhi},\ and\ \citenamefont
  {Rai}}]{Jakhad:2023ids}%
  \BibitemOpen
  \bibfield  {author} {\bibinfo {author} {\bibfnamefont {P.}~\bibnamefont
  {Jakhad}}, \bibinfo {author} {\bibfnamefont {J.}~\bibnamefont {Oudichhya}},
  \bibinfo {author} {\bibfnamefont {K.}~\bibnamefont {Gandhi}},\ and\ \bibinfo
  {author} {\bibfnamefont {A.~K.}\ \bibnamefont {Rai}},\ }\bibfield  {title}
  {\bibinfo {title} {{Identification of newly observed singly charmed baryons
  using the relativistic flux tube model}},\ }\href
  {https://doi.org/10.1103/PhysRevD.108.014011} {\bibfield  {journal} {\bibinfo
   {journal} {Phys. Rev. D}\ }\textbf {\bibinfo {volume} {108}},\ \bibinfo
  {pages} {014011} (\bibinfo {year} {2023})},\ \Eprint
  {https://arxiv.org/abs/2307.06291} {arXiv:2307.06291 [hep-ph]} \BibitemShut
  {NoStop}%
\bibitem [{\citenamefont {Jakhad}\ \emph
  {et~al.}(2024{\natexlab{a}})\citenamefont {Jakhad}, \citenamefont
  {Oudichhya},\ and\ \citenamefont {Rai}}]{Jakhad:2024fin}%
  \BibitemOpen
  \bibfield  {author} {\bibinfo {author} {\bibfnamefont {P.}~\bibnamefont
  {Jakhad}}, \bibinfo {author} {\bibfnamefont {J.}~\bibnamefont {Oudichhya}},\
  and\ \bibinfo {author} {\bibfnamefont {A.~K.}\ \bibnamefont {Rai}},\
  }\bibfield  {title} {\bibinfo {title} {{Interpretation of recently discovered
  single bottom baryons in the relativistic flux tube model}},\ }\href
  {https://doi.org/10.1103/PhysRevD.110.094005} {\bibfield  {journal} {\bibinfo
   {journal} {Phys. Rev. D}\ }\textbf {\bibinfo {volume} {110}},\ \bibinfo
  {pages} {094005} (\bibinfo {year} {2024}{\natexlab{a}})},\ \Eprint
  {https://arxiv.org/abs/2407.01655} {arXiv:2407.01655 [hep-ph]} \BibitemShut
  {NoStop}%
\bibitem [{\citenamefont {Jakhad}\ \emph
  {et~al.}(2024{\natexlab{b}})\citenamefont {Jakhad}, \citenamefont
  {Oudichhya},\ and\ \citenamefont {Rai}}]{Jakhad:2024wpx}%
  \BibitemOpen
  \bibfield  {author} {\bibinfo {author} {\bibfnamefont {P.}~\bibnamefont
  {Jakhad}}, \bibinfo {author} {\bibfnamefont {J.}~\bibnamefont {Oudichhya}},\
  and\ \bibinfo {author} {\bibfnamefont {A.~K.}\ \bibnamefont {Rai}},\
  }\bibfield  {title} {\bibinfo {title} {{Masses of higher excited states of
  \ensuremath{\Xi}c' and \ensuremath{\Omega}c baryons}},\ }\href
  {https://doi.org/10.1142/S0217751X2443005X} {\bibfield  {journal} {\bibinfo
  {journal} {Int. J. Mod. Phys. A}\ }\textbf {\bibinfo {volume} {39}},\
  \bibinfo {pages} {2443005} (\bibinfo {year} {2024}{\natexlab{b}})},\ \Eprint
  {https://arxiv.org/abs/2409.07789} {arXiv:2409.07789 [hep-ph]} \BibitemShut
  {NoStop}%
\bibitem [{\citenamefont {Jakhad}\ and\ \citenamefont
  {Rai}()}]{Private_communication}%
  \BibitemOpen
  \bibfield  {author} {\bibinfo {author} {\bibfnamefont {P.}~\bibnamefont
  {Jakhad}}\ and\ \bibinfo {author} {\bibfnamefont {A.~K.}\ \bibnamefont
  {Rai}},\ }\bibfield  {title} {\bibinfo {title} {{Relativistic flux-tube model
  predictions from charmed mesons to double-charmed baryons}},\ }\href@noop {}
  {\bibinfo  {journal} {[Communicated]}\ }\BibitemShut {NoStop}%
\bibitem [{\citenamefont {Chen}\ \emph {et~al.}(2015)\citenamefont {Chen},
  \citenamefont {Wei},\ and\ \citenamefont {Zhang}}]{article}%
  \BibitemOpen
\bibfield  {journal} {  }\bibfield  {author} {\bibinfo {author} {\bibfnamefont
  {B.}~\bibnamefont {Chen}}, \bibinfo {author} {\bibfnamefont {K.-W.}\
  \bibnamefont {Wei}},\ and\ \bibinfo {author} {\bibfnamefont {A.}~\bibnamefont
  {Zhang}},\ }\bibfield  {title} {\bibinfo {title} {Investigation of
  \ensuremath{\Lambda}$_{Q}$ and \ensuremath{\Xi}$_{Q}$ baryons in the heavy
  quark-light diquark picture},\ }\href
  {https://doi.org/10.1140/epja/i2015-15082-3} {\bibfield  {journal} {\bibinfo
  {journal} {Eur. Phys. J. A}\ }\textbf {\bibinfo {volume} {51}} (\bibinfo
  {year} {2015})}\BibitemShut {NoStop}%
\bibitem [{\citenamefont {Olson}\ \emph {et~al.}(1994)\citenamefont {Olson},
  \citenamefont {Olsson},\ and\ \citenamefont {LaCourse}}]{Olson:1993ux}%
  \BibitemOpen
  \bibfield  {author} {\bibinfo {author} {\bibfnamefont {C.}~\bibnamefont
  {Olson}}, \bibinfo {author} {\bibfnamefont {M.~G.}\ \bibnamefont {Olsson}},\
  and\ \bibinfo {author} {\bibfnamefont {D.}~\bibnamefont {LaCourse}},\
  }\bibfield  {title} {\bibinfo {title} {{The Quantized relativistic flux
  tube}},\ }\href {https://doi.org/10.1103/PhysRevD.49.4675} {\bibfield
  {journal} {\bibinfo  {journal} {Phys. Rev. D}\ }\textbf {\bibinfo {volume}
  {49}},\ \bibinfo {pages} {4675} (\bibinfo {year} {1994})}\BibitemShut
  {NoStop}%
\bibitem [{\citenamefont {Olsson}(1994)}]{Olsson:1993cn}%
  \BibitemOpen
  \bibfield  {author} {\bibinfo {author} {\bibfnamefont {M.~G.}\ \bibnamefont
  {Olsson}},\ }\bibfield  {title} {\bibinfo {title} {{The Relativistic flux
  tube model of hadrons}},\ }\href {https://doi.org/10.1007/BF02734028}
  {\bibfield  {journal} {\bibinfo  {journal} {Nuovo Cim. A}\ }\textbf {\bibinfo
  {volume} {107}},\ \bibinfo {pages} {2541} (\bibinfo {year}
  {1994})}\BibitemShut {NoStop}%
\bibitem [{\citenamefont {Allen}\ \emph {et~al.}(1999)\citenamefont {Allen},
  \citenamefont {Olsson},\ and\ \citenamefont {Veseli}}]{Allen:1999dk}%
  \BibitemOpen
  \bibfield  {author} {\bibinfo {author} {\bibfnamefont {T.~J.}\ \bibnamefont
  {Allen}}, \bibinfo {author} {\bibfnamefont {M.~G.}\ \bibnamefont {Olsson}},\
  and\ \bibinfo {author} {\bibfnamefont {S.}~\bibnamefont {Veseli}},\
  }\bibfield  {title} {\bibinfo {title} {{Curved QCD string dynamics}},\ }\href
  {https://doi.org/10.1103/PhysRevD.60.074026} {\bibfield  {journal} {\bibinfo
  {journal} {Phys. Rev. D}\ }\textbf {\bibinfo {volume} {60}},\ \bibinfo
  {pages} {074026} (\bibinfo {year} {1999})},\ \Eprint
  {https://arxiv.org/abs/hep-ph/9903222} {arXiv:hep-ph/9903222} \BibitemShut
  {NoStop}%
\bibitem [{\citenamefont {Dembinski}\ \emph {et~al.}(2021)\citenamefont
  {Dembinski}, \citenamefont {Neubert}, \citenamefont {Ruf}, \citenamefont
  {Junk}, \citenamefont {Knoepfel}, \citenamefont {Lange},\ and\ \citenamefont
  {Pivarski}}]{iminuit2021}%
  \BibitemOpen
  \bibfield  {author} {\bibinfo {author} {\bibfnamefont {H.}~\bibnamefont
  {Dembinski}}, \bibinfo {author} {\bibfnamefont {S.}~\bibnamefont {Neubert}},
  \bibinfo {author} {\bibfnamefont {T.}~\bibnamefont {Ruf}}, \bibinfo {author}
  {\bibfnamefont {T.}~\bibnamefont {Junk}}, \bibinfo {author} {\bibfnamefont
  {K.}~\bibnamefont {Knoepfel}}, \bibinfo {author} {\bibfnamefont {D.~J.}\
  \bibnamefont {Lange}},\ and\ \bibinfo {author} {\bibfnamefont
  {J.}~\bibnamefont {Pivarski}},\ }\bibfield  {title} {\bibinfo {title}
  {iminuit — a python interface to minuit},\ }\href
  {https://doi.org/10.1016/j.cpc.2021.107781} {\bibfield  {journal} {\bibinfo
  {journal} {Computer Physics Communications}\ }\textbf {\bibinfo {volume}
  {262}},\ \bibinfo {pages} {107781} (\bibinfo {year} {2021})}\BibitemShut
  {NoStop}%
\bibitem [{\citenamefont {James}\ and\ \citenamefont
  {Roos}(1975)}]{minuit1975}%
  \BibitemOpen
  \bibfield  {author} {\bibinfo {author} {\bibfnamefont {F.}~\bibnamefont
  {James}}\ and\ \bibinfo {author} {\bibfnamefont {M.}~\bibnamefont {Roos}},\
  }\bibfield  {title} {\bibinfo {title} {Minuit — a system for function
  minimization and analysis of the parameter errors and correlations},\ }\href
  {https://doi.org/10.1016/0010-4655(75)90039-9} {\bibfield  {journal}
  {\bibinfo  {journal} {Computer Physics Communications}\ }\textbf {\bibinfo
  {volume} {10}},\ \bibinfo {pages} {343} (\bibinfo {year} {1975})}\BibitemShut
  {NoStop}%
\bibitem [{\citenamefont {Aaij}\ \emph {et~al.}(2020)\citenamefont {Aaij} \emph
  {et~al.}}]{LHCb:2019epo}%
  \BibitemOpen
  \bibfield  {author} {\bibinfo {author} {\bibfnamefont {R.}~\bibnamefont
  {Aaij}} \emph {et~al.} (\bibinfo {collaboration} {LHCb}),\ }\bibfield
  {title} {\bibinfo {title} {{Precision measurement of the $\Xi_{cc}^{++}$
  mass}},\ }\href {https://doi.org/10.1007/JHEP02(2020)049} {\bibfield
  {journal} {\bibinfo  {journal} {JHEP}\ }\textbf {\bibinfo {volume} {02}},\
  \bibinfo {pages} {049}},\ \Eprint {https://arxiv.org/abs/1911.08594}
  {arXiv:1911.08594 [hep-ex]} \BibitemShut {NoStop}%
\bibitem [{\citenamefont {Hu}\ and\ \citenamefont
  {Mehen}(2006)}]{PhysRevD.73.054003}%
  \BibitemOpen
  \bibfield  {author} {\bibinfo {author} {\bibfnamefont {J.}~\bibnamefont
  {Hu}}\ and\ \bibinfo {author} {\bibfnamefont {T.}~\bibnamefont {Mehen}},\
  }\bibfield  {title} {\bibinfo {title} {Chiral lagrangian with heavy
  quark-diquark symmetry},\ }\href {https://doi.org/10.1103/PhysRevD.73.054003}
  {\bibfield  {journal} {\bibinfo  {journal} {Phys. Rev. D}\ }\textbf {\bibinfo
  {volume} {73}},\ \bibinfo {pages} {054003} (\bibinfo {year}
  {2006})}\BibitemShut {NoStop}%
\bibitem [{\citenamefont {Song}\ \emph {et~al.}(2023)\citenamefont {Song},
  \citenamefont {Jia}, \citenamefont {Zhang},\ and\ \citenamefont
  {Hosaka}}]{jia2023}%
  \BibitemOpen
  \bibfield  {author} {\bibinfo {author} {\bibfnamefont {Y.}~\bibnamefont
  {Song}}, \bibinfo {author} {\bibfnamefont {D.}~\bibnamefont {Jia}}, \bibinfo
  {author} {\bibfnamefont {W.}~\bibnamefont {Zhang}},\ and\ \bibinfo {author}
  {\bibfnamefont {A.}~\bibnamefont {Hosaka}},\ }\bibfield  {title} {\bibinfo
  {title} {Low-lying doubly heavy baryons: Regge relation and mass scaling},\
  }\href {https://doi.org/10.1140/epjc/s10052-022-11136-9} {\bibfield
  {journal} {\bibinfo  {journal} {European Physical Journal C}\ }\textbf
  {\bibinfo {volume} {83}},\ \bibinfo {pages} {1} (\bibinfo {year}
  {2023})}\BibitemShut {NoStop}%
\bibitem [{\citenamefont {Chen}\ \emph {et~al.}(2020)\citenamefont {Chen},
  \citenamefont {Zhang},\ and\ \citenamefont {He}}]{PhysRevD.101.014020}%
  \BibitemOpen
  \bibfield  {author} {\bibinfo {author} {\bibfnamefont {B.}~\bibnamefont
  {Chen}}, \bibinfo {author} {\bibfnamefont {A.}~\bibnamefont {Zhang}},\ and\
  \bibinfo {author} {\bibfnamefont {J.}~\bibnamefont {He}},\ }\bibfield
  {title} {\bibinfo {title} {Bottomonium spectrum in the relativistic flux tube
  model},\ }\href {https://doi.org/10.1103/PhysRevD.101.014020} {\bibfield
  {journal} {\bibinfo  {journal} {Phys. Rev. D}\ }\textbf {\bibinfo {volume}
  {101}},\ \bibinfo {pages} {014020} (\bibinfo {year} {2020})}\BibitemShut
  {NoStop}%
\bibitem [{\citenamefont {Johnson}\ and\ \citenamefont
  {Thorn}(1976)}]{PhysRevD.13.1934}%
  \BibitemOpen
  \bibfield  {author} {\bibinfo {author} {\bibfnamefont {K.}~\bibnamefont
  {Johnson}}\ and\ \bibinfo {author} {\bibfnamefont {C.~B.}\ \bibnamefont
  {Thorn}},\ }\bibfield  {title} {\bibinfo {title} {Stringlike solutions of the
  bag model},\ }\href {https://doi.org/10.1103/PhysRevD.13.1934} {\bibfield
  {journal} {\bibinfo  {journal} {Phys. Rev. D}\ }\textbf {\bibinfo {volume}
  {13}},\ \bibinfo {pages} {1934} (\bibinfo {year} {1976})}\BibitemShut
  {NoStop}%
\bibitem [{\citenamefont {Lahde}\ \emph {et~al.}(2000)\citenamefont {Lahde},
  \citenamefont {Nyfalt},\ and\ \citenamefont {Riska}}]{Lahde:1999ih}%
  \BibitemOpen
  \bibfield  {author} {\bibinfo {author} {\bibfnamefont {T.~A.}\ \bibnamefont
  {Lahde}}, \bibinfo {author} {\bibfnamefont {C.~J.}\ \bibnamefont {Nyfalt}},\
  and\ \bibinfo {author} {\bibfnamefont {D.~O.}\ \bibnamefont {Riska}},\
  }\bibfield  {title} {\bibinfo {title} {{Spectra and M1 decay widths of heavy
  light mesons}},\ }\href {https://doi.org/10.1016/S0375-9474(00)00154-8}
  {\bibfield  {journal} {\bibinfo  {journal} {Nucl. Phys. A}\ }\textbf
  {\bibinfo {volume} {674}},\ \bibinfo {pages} {141} (\bibinfo {year}
  {2000})},\ \Eprint {https://arxiv.org/abs/hep-ph/9908485}
  {arXiv:hep-ph/9908485} \BibitemShut {NoStop}%
\bibitem [{\citenamefont {L\"u}\ \emph {et~al.}(2016)\citenamefont {L\"u},
  \citenamefont {Pan}, \citenamefont {Wang}, \citenamefont {Wang},\ and\
  \citenamefont {Li}}]{Lu:2016bbk}%
  \BibitemOpen
  \bibfield  {author} {\bibinfo {author} {\bibfnamefont {Q.-F.}\ \bibnamefont
  {L\"u}}, \bibinfo {author} {\bibfnamefont {T.-T.}\ \bibnamefont {Pan}},
  \bibinfo {author} {\bibfnamefont {Y.-Y.}\ \bibnamefont {Wang}}, \bibinfo
  {author} {\bibfnamefont {E.}~\bibnamefont {Wang}},\ and\ \bibinfo {author}
  {\bibfnamefont {D.-M.}\ \bibnamefont {Li}},\ }\bibfield  {title} {\bibinfo
  {title} {{Excited bottom and bottom-strange mesons in the quark model}},\
  }\href {https://doi.org/10.1103/PhysRevD.94.074012} {\bibfield  {journal}
  {\bibinfo  {journal} {Phys. Rev. D}\ }\textbf {\bibinfo {volume} {94}},\
  \bibinfo {pages} {074012} (\bibinfo {year} {2016})},\ \Eprint
  {https://arxiv.org/abs/1607.02812} {arXiv:1607.02812 [hep-ph]} \BibitemShut
  {NoStop}%
\bibitem [{\citenamefont {Akers}\ \emph {et~al.}(1995)\citenamefont {Akers}
  \emph {et~al.}}]{OPAL:1994hqv}%
  \BibitemOpen
  \bibfield  {author} {\bibinfo {author} {\bibfnamefont {R.}~\bibnamefont
  {Akers}} \emph {et~al.} (\bibinfo {collaboration} {OPAL}),\ }\bibfield
  {title} {\bibinfo {title} {{Observations of pi - B charge - flavor
  correlations and resonant B pi and B K production}},\ }\href
  {https://doi.org/10.1007/BF01496577} {\bibfield  {journal} {\bibinfo
  {journal} {Z. Phys. C}\ }\textbf {\bibinfo {volume} {66}},\ \bibinfo {pages}
  {19} (\bibinfo {year} {1995})}\BibitemShut {NoStop}%
\bibitem [{\citenamefont {Abreu}\ \emph {et~al.}(1995)\citenamefont {Abreu}
  \emph {et~al.}}]{DELPHI:1994fnu}%
  \BibitemOpen
  \bibfield  {author} {\bibinfo {author} {\bibfnamefont {P.}~\bibnamefont
  {Abreu}} \emph {et~al.} (\bibinfo {collaboration} {DELPHI}),\ }\bibfield
  {title} {\bibinfo {title} {{Observation of orbitally excited B mesons}},\
  }\href {https://doi.org/10.1016/0370-2693(94)01696-A} {\bibfield  {journal}
  {\bibinfo  {journal} {Phys. Lett. B}\ }\textbf {\bibinfo {volume} {345}},\
  \bibinfo {pages} {598} (\bibinfo {year} {1995})}\BibitemShut {NoStop}%
\bibitem [{\citenamefont {Barate}\ \emph {et~al.}(1998)\citenamefont {Barate}
  \emph {et~al.}}]{ALEPH:1998unp}%
  \BibitemOpen
  \bibfield  {author} {\bibinfo {author} {\bibfnamefont {R.}~\bibnamefont
  {Barate}} \emph {et~al.} (\bibinfo {collaboration} {ALEPH}),\ }\bibfield
  {title} {\bibinfo {title} {{Resonant structure and flavor tagging in the B
  pi+- system using fully reconstructed B decays}},\ }\href
  {https://doi.org/10.1016/S0370-2693(98)00180-4} {\bibfield  {journal}
  {\bibinfo  {journal} {Phys. Lett. B}\ }\textbf {\bibinfo {volume} {425}},\
  \bibinfo {pages} {215} (\bibinfo {year} {1998})}\BibitemShut {NoStop}%
\bibitem [{\citenamefont {Acciarri}\ \emph {et~al.}(1999)\citenamefont
  {Acciarri} \emph {et~al.}}]{L3:1999pdo}%
  \BibitemOpen
  \bibfield  {author} {\bibinfo {author} {\bibfnamefont {M.}~\bibnamefont
  {Acciarri}} \emph {et~al.} (\bibinfo {collaboration} {L3}),\ }\bibfield
  {title} {\bibinfo {title} {{Measurement of the spectroscopy of orbitally
  excited $B$ mesons at LEP}},\ }\href
  {https://doi.org/10.1016/S0370-2693(99)01067-9} {\bibfield  {journal}
  {\bibinfo  {journal} {Phys. Lett. B}\ }\textbf {\bibinfo {volume} {465}},\
  \bibinfo {pages} {323} (\bibinfo {year} {1999})},\ \Eprint
  {https://arxiv.org/abs/hep-ex/9909018} {arXiv:hep-ex/9909018} \BibitemShut
  {NoStop}%
\bibitem [{\citenamefont {Xiao}\ and\ \citenamefont
  {Zhong}(2014)}]{Xiao:2014ura}%
  \BibitemOpen
  \bibfield  {author} {\bibinfo {author} {\bibfnamefont {L.-Y.}\ \bibnamefont
  {Xiao}}\ and\ \bibinfo {author} {\bibfnamefont {X.-H.}\ \bibnamefont
  {Zhong}},\ }\bibfield  {title} {\bibinfo {title} {{Strong decays of higher
  excited heavy-light mesons in a chiral quark model}},\ }\href
  {https://doi.org/10.1103/PhysRevD.90.074029} {\bibfield  {journal} {\bibinfo
  {journal} {Phys. Rev. D}\ }\textbf {\bibinfo {volume} {90}},\ \bibinfo
  {pages} {074029} (\bibinfo {year} {2014})},\ \Eprint
  {https://arxiv.org/abs/1407.7408} {arXiv:1407.7408 [hep-ph]} \BibitemShut
  {NoStop}%
\bibitem [{\citenamefont {Yu}\ and\ \citenamefont {Wang}(2020)}]{Yu:2019iwm}%
  \BibitemOpen
  \bibfield  {author} {\bibinfo {author} {\bibfnamefont {G.-L.}\ \bibnamefont
  {Yu}}\ and\ \bibinfo {author} {\bibfnamefont {Z.-G.}\ \bibnamefont {Wang}},\
  }\bibfield  {title} {\bibinfo {title} {{Analysis of the excited bottom and
  bottom-strange states $B_{1}(5721)$, $B_{2}^{*}(5747)$, $B_{s1}(5830)$,
  $B_{s2}^{*}(5840)$, $B_{J}(5840)$ and $B_{J}(5970)$ in $B$ meson family}},\
  }\href {https://doi.org/10.1088/1674-1137/44/3/033103} {\bibfield  {journal}
  {\bibinfo  {journal} {Chin. Phys. C}\ }\textbf {\bibinfo {volume} {44}},\
  \bibinfo {pages} {033103} (\bibinfo {year} {2020})},\ \Eprint
  {https://arxiv.org/abs/1907.08760} {arXiv:1907.08760 [hep-ph]} \BibitemShut
  {NoStop}%
\bibitem [{\citenamefont {Asghar}\ \emph {et~al.}(2018)\citenamefont {Asghar},
  \citenamefont {Masud}, \citenamefont {Swanson}, \citenamefont {Akram},\ and\
  \citenamefont {Atif~Sultan}}]{Asghar:2018tha}%
  \BibitemOpen
  \bibfield  {author} {\bibinfo {author} {\bibfnamefont {I.}~\bibnamefont
  {Asghar}}, \bibinfo {author} {\bibfnamefont {B.}~\bibnamefont {Masud}},
  \bibinfo {author} {\bibfnamefont {E.~S.}\ \bibnamefont {Swanson}}, \bibinfo
  {author} {\bibfnamefont {F.}~\bibnamefont {Akram}},\ and\ \bibinfo {author}
  {\bibfnamefont {M.}~\bibnamefont {Atif~Sultan}},\ }\bibfield  {title}
  {\bibinfo {title} {{Decays and spectrum of bottom and bottom strange
  mesons}},\ }\href {https://doi.org/10.1140/epja/i2018-12558-6} {\bibfield
  {journal} {\bibinfo  {journal} {Eur. Phys. J. A}\ }\textbf {\bibinfo {volume}
  {54}},\ \bibinfo {pages} {127} (\bibinfo {year} {2018})},\ \Eprint
  {https://arxiv.org/abs/1804.08802} {arXiv:1804.08802 [hep-ph]} \BibitemShut
  {NoStop}%
\bibitem [{\citenamefont {Hao}\ \emph {et~al.}(2023)\citenamefont {Hao},
  \citenamefont {Lu},\ and\ \citenamefont {Wang}}]{Hao:2022ibj}%
  \BibitemOpen
  \bibfield  {author} {\bibinfo {author} {\bibfnamefont {W.}~\bibnamefont
  {Hao}}, \bibinfo {author} {\bibfnamefont {Y.}~\bibnamefont {Lu}},\ and\
  \bibinfo {author} {\bibfnamefont {E.}~\bibnamefont {Wang}},\ }\bibfield
  {title} {\bibinfo {title} {{The assignments of the $B_s$ mesons within the
  screened potential model and $^3P_0$ model}},\ }\href
  {https://doi.org/10.1140/epjc/s10052-023-11689-3} {\bibfield  {journal}
  {\bibinfo  {journal} {Eur. Phys. J. C}\ }\textbf {\bibinfo {volume} {83}},\
  \bibinfo {pages} {520} (\bibinfo {year} {2023})},\ \Eprint
  {https://arxiv.org/abs/2212.10068} {arXiv:2212.10068 [hep-ph]} \BibitemShut
  {NoStop}%
\bibitem [{\citenamefont {Karliner}\ and\ \citenamefont
  {Rosner}(2017)}]{Karliner:2017kfm}%
  \BibitemOpen
  \bibfield  {author} {\bibinfo {author} {\bibfnamefont {M.}~\bibnamefont
  {Karliner}}\ and\ \bibinfo {author} {\bibfnamefont {J.~L.}\ \bibnamefont
  {Rosner}},\ }\bibfield  {title} {\bibinfo {title} {{Very narrow excited
  $\Omega_c$ baryons}},\ }\href {https://doi.org/10.1103/PhysRevD.95.114012}
  {\bibfield  {journal} {\bibinfo  {journal} {Phys. Rev. D}\ }\textbf {\bibinfo
  {volume} {95}},\ \bibinfo {pages} {114012} (\bibinfo {year} {2017})},\
  \Eprint {https://arxiv.org/abs/1703.07774} {arXiv:1703.07774 [hep-ph]}
  \BibitemShut {NoStop}%
\end{thebibliography}%
\end{document}